\documentclass[ALICE,manyauthors]{cernphprep}
\usepackage[comma,square,numbers,sort&compress]{natbib}

\usepackage{lineno}
\usepackage{xspace}
\usepackage{hyperref}
\usepackage{booktabs}
\usepackage[T1]{fontenc}
\usepackage{orcidlink}

\begin{document}
%

\newcommand{\pp}           {pp\xspace}
\newcommand{\ppbar}        {\mbox{$\mathrm {p\overline{p}}$}\xspace}
\newcommand{\XeXe}         {\mbox{Xe--Xe}\xspace}
\newcommand{\PbPb}         {\mbox{Pb--Pb}\xspace}
\newcommand{\pA}           {\mbox{pA}\xspace}
\newcommand{\pPb}          {\mbox{p--Pb}\xspace}
\newcommand{\AuAu}         {\mbox{Au--Au}\xspace}
\newcommand{\dAu}          {\mbox{d--Au}\xspace}
\newcommand{\kt}{\ensuremath{k_{\rm T}}\xspace}
\newcommand{\nbinv}{\ensuremath{\rm nb^{-1}}}
\newcommand {\ubinv}{\ensuremath{\mu\rm b^{-1}}}
\newcommand{\mubinv}{\ubinv}
\newcommand {\um}{\ensuremath{\mu\rm m}\xspace}
\newcommand{\mub}{\ensuremath{\mu\rm b}\xspace}
\newcommand{\mb}{\ensuremath{\rm mb}\xspace}
\newcommand{\s}            {\ensuremath{\sqrt{s}}\xspace}
\newcommand{\snn}          {\ensuremath{\sqrt{s_{\mathrm{NN}}}}\xspace}
\newcommand{\pt}           {\ensuremath{p_{\rm T}}\xspace}
\newcommand{\meanpt}       {$\langle p_{\mathrm{T}}\rangle$\xspace}
\newcommand{\ycms}         {\ensuremath{y_{\rm CMS}}\xspace}
\newcommand{\ylab}         {\ensuremath{y_{\rm lab}}\xspace}
\newcommand{\etarange}[1]  {\mbox{$\left | \eta \right |~<~#1$}}
\newcommand{\yrange}[1]    {\mbox{$\left | y \right |~<~#1$}}
\newcommand{\dndy}         {\ensuremath{\mathrm{d}N_\mathrm{ch}/\mathrm{d}y}\xspace}
\newcommand{\dndeta}       {\ensuremath{\mathrm{d}N_\mathrm{ch}/\mathrm{d}\eta}\xspace}
\newcommand{\avdndeta}     {\ensuremath{\langle\dndeta\rangle}\xspace}
\newcommand{\dNdy}         {\ensuremath{\mathrm{d}N_\mathrm{ch}/\mathrm{d}y}\xspace}
\newcommand{\Npart}        {\ensuremath{N_\mathrm{part}}\xspace}
\newcommand{\Ncoll}        {\ensuremath{N_\mathrm{coll}}\xspace}
\newcommand{\dEdx}         {\ensuremath{\textrm{d}E/\textrm{d}x}\xspace}
\newcommand{\RpPb}         {\ensuremath{R_{\rm pPb}}\xspace}

\newcommand{\lumi}         {\ensuremath{\mathcal{L}}\xspace}

\newcommand{\Lint}         {\ensuremath{\mathcal{L}_\mathrm{int}}\xspace}

\newcommand{\nineH}        {$\sqrt{s}~=~0.9$~Te\kern-.1emV\xspace}
\newcommand{\seven}        {$\sqrt{s}~=~7$~Te\kern-.1emV\xspace}
\newcommand{\twoH}         {$\sqrt{s}~=~0.2$~Te\kern-.1emV\xspace}
\newcommand{\twosevensix}  {$\sqrt{s}~=~2.76$~Te\kern-.1emV\xspace}
\newcommand{\five}         {$\sqrt{s}~=~5.02$~Te\kern-.1emV\xspace}
\newcommand{\thirteen}     {$\sqrt{s}~=~13$~Te\kern-.1emV\xspace}
\newcommand{\twosevensixnn}{$\sqrt{s_{\mathrm{NN}}}~=~2.76$~Te\kern-.1emV\xspace}
\newcommand{\fivenn}       {$\sqrt{s_{\mathrm{NN}}}~=~5.02$~Te\kern-.1emV\xspace}
\newcommand{\eightnn}       {$\sqrt{s_{\mathrm{NN}}}~=~8.16$~Te\kern-.1emV\xspace}
\newcommand{\LT}           {L{\'e}vy-Tsallis\xspace}
\newcommand{\GeVc}         {Ge\kern-.1emV/$c$\xspace}
\newcommand{\MeVc}         {Me\kern-.1emV/$c$\xspace}
\newcommand{\TeV}          {Te\kern-.1emV\xspace}
\newcommand{\GeV}          {Ge\kern-.1emV\xspace}
\newcommand{\MeV}          {Me\kern-.1emV\xspace}
\newcommand{\GeVmass}      {Ge\kern-.1emV/$c^2$\xspace}
\newcommand{\MeVmass}      {Me\kern-.1emV/$c^2$\xspace}

\newcommand {\ep}        {\mbox{$\mathrm {e^-p}$}\xspace}

\newcommand{\ITS}          {\rm{ITS}\xspace}
\newcommand{\TOF}          {\rm{TOF}\xspace}
\newcommand{\ZDC}          {\rm{ZDC}\xspace}
\newcommand{\ZDCs}         {\rm{ZDCs}\xspace}
\newcommand{\ZNA}          {\rm{ZNA}\xspace}
\newcommand{\ZNC}          {\rm{ZNC}\xspace}
\newcommand{\SPD}          {\rm{SPD}\xspace}
\newcommand{\SDD}          {\rm{SDD}\xspace}
\newcommand{\SSD}          {\rm{SSD}\xspace}
\newcommand{\TPC}          {\rm{TPC}\xspace}
\newcommand{\TRD}          {\rm{TRD}\xspace}
\newcommand{\VZERO}        {\rm{V0}\xspace}
\newcommand{\VZEROA}       {\rm{V0A}\xspace}
\newcommand{\VZEROC}       {\rm{V0C}\xspace}
\newcommand{\Vdecay} 	   {\ensuremath{V^{0}}\xspace}

\newcommand{\ee}           {\ensuremath{\rm e^{+}e^{-}}} 
\newcommand{\pip}          {\ensuremath{\pi^{+}}\xspace}
\newcommand{\pim}          {\ensuremath{\pi^{-}}\xspace}
\newcommand{\kap}          {\ensuremath{\rm{K}^{+}}\xspace}
\newcommand{\kam}          {\ensuremath{\rm{K}^{-}}\xspace}
\newcommand{\pbar}         {\ensuremath{\rm\overline{p}}\xspace}
\newcommand{\kzero}        {\ensuremath{{\rm K}^{0}_{\rm{S}}}\xspace}
\newcommand{\lmb}          {\ensuremath{\Lambda}\xspace}
\newcommand{\almb}         {\ensuremath{\overline{\Lambda}}\xspace}
\newcommand{\Om}           {\ensuremath{\Omega^-}\xspace}
\newcommand{\Mo}           {\ensuremath{\overline{\Omega}^+}\xspace}
\newcommand{\X}            {\ensuremath{\Xi^-}\xspace}
\newcommand{\Ix}           {\ensuremath{\overline{\Xi}^+}\xspace}
\newcommand{\Xis}          {\ensuremath{\Xi^{\pm}}\xspace}
\newcommand{\Oms}          {\ensuremath{\Omega^{\pm}}\xspace}
\newcommand{\degree}       {\ensuremath{^{\rm o}}\xspace}

\newcommand{\proton}    {\mbox{$\mathrm {p}$}\xspace}
\newcommand{\pOuPbar}   {\mbox{$\mathrm {p^{\pm}}$}\xspace}
\newcommand{\DZero}     {\ensuremath{\mathrm {D^0}}\xspace}
\newcommand{\DZerobar}  {\ensuremath{\mathrm {\overline{D}^0}}\xspace}
\newcommand{\Bminus}    {\ensuremath{\mathrm {B^-}}\xspace}
\newcommand{\BZero}     {\ensuremath{\mathrm {B^0}}\xspace}
\newcommand{\BZerobar}  {\ensuremath{\mathrm {\overline{B}{}^0}}\xspace}
\newcommand{\Bs}     {\ensuremath{\mathrm {B^0_s}}\xspace}
\newcommand{\Bsbar}  {\ensuremath{\mathrm {\overline{B}{}^0_s}}\xspace}

\newcommand{\Dmes}       {\ensuremath{\mathrm {D}}\xspace}

\newcommand{\Lb}{\ensuremath{\rm {\Lambda_b^{0}}}\xspace}
\newcommand{\Xic}         {\ensuremath{\mathrm {\Xi_{c}}}\xspace}
\newcommand{\lambdab}     {\ensuremath{\mathrm {\Lambda_{b}^{0}}}\xspace}
\newcommand{\lambdac}     {\ensuremath{\mathrm {\Lambda_{c}^{+}}}\xspace}
\newcommand{\xicz}        {\ensuremath{\mathrm {\Xi_{c}^{0}}}\xspace}
\newcommand{\xiczp}        {\ensuremath{\mathrm {\Xi_{c}^{0,+}}}\xspace}
\newcommand{\XicD} {\ensuremath{\xiczp/\Dz}\xspace}
\newcommand{\xicp}        {\ensuremath{\mathrm {\Xi_{c}^{+}}}\xspace}
\newcommand{\xib}        {\ensuremath{\mathrm {\Xi_{b}}}\xspace}
\newcommand{\LambdaParticle}        {\ensuremath{\mathrm {\Lambda}}\xspace}

\newcommand{\rmLambdaZ}         {\ensuremath{\mathrm {\Lambda}}\xspace}
\newcommand{\rmAlambdaZ}        {\ensuremath{\mathrm {\overline{\Lambda}}}\xspace}
\newcommand{\rmLambda}          {\ensuremath{\mathrm {\Lambda}}\xspace}
\newcommand{\rmAlambda}         {\ensuremath{\mathrm {\overline{\Lambda}}}\xspace}
\newcommand{\rmLambdas}         {\ensuremath{\mathrm {\Lambda \kern-0.2em + \kern-0.2em \overline{\Lambda}}}\xspace}

\newcommand{\Vzero}             {\ensuremath{\mathrm {V^0}}\xspace}
\newcommand{\Vzerob}             {\ensuremath{{\bold \mathrm {V^0}}}\xspace}
\newcommand{\Kzero}             {\ensuremath{\mathrm {K^0}}\xspace}
\newcommand{\Kzs}               {\ensuremath{\mathrm {K^0_S}}\xspace}
\newcommand{\phimes}            {\ensuremath{\mathrm {\phi}}\xspace}
\newcommand{\Kminus}            {\ensuremath{\mathrm {K^-}}\xspace}
\newcommand{\Kplus}             {\ensuremath{\mathrm {K^+}}\xspace}
\newcommand{\Kstar}             {\ensuremath{\mathrm {K^*}}\xspace}
\newcommand{\Kplusmin}          {\ensuremath{\mathrm {K^{\pm}}}\xspace}
\newcommand{\Jpsi}              {\ensuremath{\rm J/\psi}\xspace}
\newcommand{\DtoKpi}{\ensuremath{\rm D^0\to K^-\pi^+}\xspace}
\newcommand{\DtoKpipi}{\ensuremath{\rm D^+\to K^-\pi^+\pi^+}\xspace}
\newcommand{\DstartoDpi}{\ensuremath{\rm D^{*+}\to D^0\pi^+}\xspace}
\newcommand{\Dzero}{\ensuremath{\mathrm {D^0}}\xspace}
\newcommand{\Dz}{\Dzero}
\newcommand{\Dzerobar}{\ensuremath{\mathrm{\overline{D}^0}}\xspace}
\newcommand{\Dstar}{\ensuremath{\rm D^{*+}}\xspace}
\newcommand{\Dplus}{\ensuremath{\rm D^+}\xspace}
\newcommand{\Dp}{\Dplus}
\newcommand{\Dsubs}{\ensuremath{\rm D_{s}^+}\xspace}
\newcommand{\Ds}{\Dsubs}
\newcommand{\decleng}{\ensuremath{\rm L}_{xyz}}
\newcommand{\Lcminus}{\ensuremath{\rm {\overline{\Lambda}{}_c^-}}\xspace}
\newcommand{\Lcplus}{\lambdac}
\newcommand{\Lc}         {\Lcplus}
\newcommand{\LcD} {\ensuremath{\lambdac/\Dzero}\xspace}

\newcommand{\xiczDz} {\ensuremath{\xicz/\Dzero}\xspace}

\newcommand{\Lbzero}{\ensuremath{\rm {\Lambda_b^0}}\xspace}
\newcommand{\LctopKpi}{\ensuremath{\rm \Lambda_{c}^{+}\to p K^-\pi^+}\xspace}
\newcommand{\LcpKpi}{\LctopKpi}
\newcommand{\LbtoLc}{\ensuremath{\rm \Lambda_{b}^{0}\to \Lc + \rm{X}}\xspace}
\newcommand{\LctopKzS}{\ensuremath{\rm \Lambda_{c}^{+}\to p K^{0}_{S}}\xspace}
\newcommand{\LcpKs}{\LctopKzS}
\newcommand{\LctopKs}{\LctopKzS}
\newcommand{\LctoenuLambda}{\ensuremath{\rm \Lambda_{c}^{+}\to e^{+} \nu_{e} \Lambda}\xspace}
\newcommand{\cosP}{\ensuremath{\rm cos_{\Theta_{pointing}}}\xspace}
\newcommand{\KzStopippim}{\ensuremath{\rm K^{0}_{S}\to \pi^{+} \pi^{-}}\xspace}
\newcommand{\Lambdatoppim}{\ensuremath{\rm \Lambda \to p \pi^{-}}\xspace}
\newcommand{\nue}{$\nu_e$}
\newcommand{\DtopiKzs}{\ensuremath{\rm D^+\to \pi^+ K^{0}_{S}}\xspace}
\newcommand{\DstoKKzs}{\ensuremath{\rm D_s^+\to K^+ K^{0}_{S}}\xspace}

\newcommand{\ptLc}{\ensuremath{p_{\rm T, \Lambda_c}}\xspace}
\newcommand{\ptpion}{\ensuremath{p_{\rm T, \pi}}\xspace}
\newcommand{\ptK}{\ensuremath{p_{\rm T, K}}\xspace}
\newcommand{\ptproton}{\ensuremath{p_{\rm T, \proton}}\xspace}
\newcommand{\Omegac}{\ensuremath{\Omega_{\rm c}^{0}}\xspace}
\newcommand{\Sigmac}{\ensuremath{\Sigma_{\rm c}^{0,++}}\xspace}

\begin{titlepage}
\PHyear{2024}      
\PHnumber{140}     
\PHdate{22 May} 

\title{Measurement of the production cross section of prompt $\Xi^0_{\rm c}$ baryons in p--Pb collisions at $\mathbf{\sqrt{s_{\mathrm{\textbf{NN}}}}~=~5.02}$~Te\kern-.1emV\xspace}

\ShortTitle{Prompt $\Xi^0_{\rm c}$-baryon production in p--Pb collisions at \fivenn}   

\Collaboration{ALICE Collaboration\thanks{See Appendix~\ref{app:collab} for the list of collaboration members}}
\ShortAuthor{ALICE Collaboration} 

\begin{abstract}
The transverse momentum (\pt) differential production cross section of the promptly produced charm-strange baryon \xicz (and its charge conjugate $\overline{\xicz}$) is measured at midrapidity via its hadronic decay into ${\rm \pi^{+}}\Xi^{-}$ in p--Pb collisions at a centre-of-mass energy per nucleon--nucleon collision \fivenn with the ALICE detector at the LHC. The \xicz nuclear modification factor (\RpPb), calculated from the cross sections in pp and p--Pb collisions, is presented and compared with the \RpPb of \Lc baryons. The ratios between the \pt-differential production cross section of \xicz baryons and those of \Dzero mesons and \Lc baryons are also reported and compared with results at forward and backward rapidity from the LHCb Collaboration. The measurements of the production cross section of prompt $\Xi^0_{\rm c}$ baryons are compared with a model based on perturbative QCD calculations of charm-quark production cross sections, which includes only cold nuclear matter effects in \pPb collisions, and underestimates the measurement by a factor of about 50.  This discrepancy is reduced when the data is compared with a model that includes string formation beyond leading-colour approximation or in which hadronisation is implemented via quark coalescence. The \pt-integrated cross section of prompt $\Xi^0_{\rm c}$-baryon production at midrapidity extrapolated down to \pt = 0 is also reported. These measurements offer insights and constraints for theoretical calculations of the hadronisation process. Additionally, they provide inputs for the calculation of the charm production cross section in \pPb collisions at midrapidity. 
\end{abstract}
\end{titlepage}

\setcounter{page}{2}


\section{Introduction} \label{sec:intro}

Measurements of heavy-flavour hadron production in hadronic collisions provide crucial tests for calculations based on quantum chromodynamics (QCD). Calculations of \pt-differential heavy-flavour hadron production cross sections in hadronic collisions are factorised into three separate components: the parton distribution functions (PDFs), which describe the Bjorken-$x$ distributions of quarks and gluons within the incoming hadrons for a given transferred momentum squared $Q^2$; the hard-scattering cross section for the partons to produce a charm- or beauty-quark pair; and the fragmentation functions, which characterise the hadronisation of a quark to a given hadron species~\cite{Collins:1989gx}. The hadronisation process is typically described via two different mechanisms: fragmentation and recombination (also known as coalescence). In the former, colour neutrality is reached by means of gluon radiation and gluon splitting into quark pairs, with a final phase in which quark pairs or triplets bind together to form mesons or baryons. In the latter, two or three quarks close in the velocity and coordinate space, bind to form color-neutral hadrons. As charm and beauty quarks have masses much larger than $\Lambda_\mathrm{QCD}$, which is the energy scale in QCD at which quarks and gluons are confined within hadrons, the parton--parton hard-scattering cross sections can be calculated perturbatively~\cite{Cacciari:1998it}. In contrast, the fragmentation functions cannot be calculated with perturbative QCD (pQCD) methods, so they are determined from measurements in \ee collisions. They are then applied in cross section calculations, assuming that the relevant hadronisation processes are ``universal'', i.e.~independent of the collision energy and system. A recent review with a more comprehensive overview of heavy-quark hadronisation can be found in Ref.~\cite{Altmann:2024kwx}. 
Factorisation can be implemented in pQCD-based calculations in different ways, for example, in terms of the transferred momentum squared (collinear factorisation)~\cite{Collins:1989gx}. Calculations for LHC energies, like the general-mass variable-flavour-number scheme (GM-VFNS)~\cite{Kramer:2017gct, Helenius:2018uul} and the fixed order plus next-to-leading logarithms (FONLL) approach~\cite{fonllcalc1, fonllcalc2} provide a next-to-leading order (NLO) accuracy with all-order resummation of next-to-leading logarithms. These calculations describe the production of heavy-flavour mesons within uncertainties in wide kinematic and pp collision-energy ranges~\cite{ALICE:2022wpn, CMS:2021lab, LHCb:2016ikn, LHCb:2015swx}. The dominant source of theoretical uncertainty in these calculations is related to the choice of the energy scales for the validity of the perturbative regime (factorisation and renormalisation scales). 
Charm-hadron production cross sections are also compared with the POWHEG pQCD calculations of charm-quark production with NLO accuracy~\cite{Frixione:2007nw}, matched with PYTHIA~6~\cite{sjostrand2006pythia} to generate the parton shower and fragmentation. The POWHEG+PYTHIA~6 simulations describe the charm-meson production cross sections but largely underpredict the production of charm baryons~\cite{ALICE:2022exq}. To isolate the effects of hadronisation, hadron-to-hadron production ratios within the charm sector, such as $\Dsubs/\Dzero$, \LcD, and \xiczDz are particularly effective, since in pQCD calculations the PDFs and the choice of the factorisation and renormalisation scales are common to all charm-hadron species and their effects almost fully cancel in the yield ratios.  

Previous measurements of charm-meson production cross sections in pp and p--Pb collisions at the LHC~\cite{ALICE:2021mgk, ALICE:2017olh, LHCb:2016ikn, LHCb:2023rpm, CMS:2021lab, ALICE:2023sgl, ALICE:2016yta} show that the $\Dplus/\Dz$ ratio is independent of the transverse momentum within uncertainties, while a hint of increase with \pt is visible for the $\Ds/\Dz$, $\Ds/\Dplus$, and $\Ds/(\Dz + \Dplus)$ ratios in the interval \pt $<$ 8 GeV$/c$.
The ratios are also described well by pQCD calculations and by the PYTHIA~8 event generator using the Monash tune~\cite{Sjostrand:2014zea,Skands:2014pea}, with the fragmentation tuned on \ee collisions. However, the charm baryon-to-meson ratios \LcD, \XicD, $\Omegac/\Dz$, and $\Sigmac/\Dz$ measured at midrapidity at the LHC~\cite{ALICE:2017thy, ALICE:2020wla,  ALICE:2020wfu,CMS:2019uws,ALICE:2021psx, ALICE:2021bli, ALICE:2021rzj, ALICE:OmegaC, ALICE:2022exq, LHCb:2018weo} show significant deviations from the values measured in \ee collisions, and the Monash tune of PYTHIA~8 significantly underpredicts the production rates of charm baryons. These results pose a challenge to the assumption of a universal hadronisation mechanism~\cite{Dai:2024vjy}. Models that incorporate other effects such as string formation beyond the leading-colour approximation~\cite{Christiansen:2015yqa,Bierlich:2023okq}, coalescence or recombination of charm quarks with quarks or di-quarks from a thermal medium~\cite{Minissale:2020bif,Song:2018tpv, Beraudo:2023nlq}, and statistical hadronisation including higher charm-resonant states not yet discovered~\cite{He:2019tik} provide a better description of the data. Measurements of beauty-baryon production in pp and p--Pb collisions by the CMS, LHCb, and ALICE Collaborations~\cite{CMS:2012wje,LHCb:2019fns,LHCb:2015qvk, LHCb:2019avm, LHCb:2023wbo, ALICE:2023wbx} also indicate similar differences in hadronisation mechanisms in the beauty sector between hadronic and leptonic collision systems~\cite{Dai:2024vjy,Bierlich:2023okq,Christiansen:2015yqa}. The enhancement of the relative abundance of baryons compared to that of mesons has a strong impact on the charm-quark fragmentation fractions, as demonstrated in pp collisions at $\mbox{\s = 5.02}$~\TeV and \s = 13 \TeV ~\cite{ALICE:2021dhb,ALICE:2023sgl}.

Measurements in proton--nucleus collisions allow an assessment of the various effects, denoted as cold-nuclear-matter (CNM) effects, related to
the presence of one or more nuclei in the colliding system.
In the initial state, the PDFs are modified in bound nucleons compared to free nucleons, depending on $x$ and $Q^2$~\cite{Arneodo:1992wf,Malace:2014uea}.
At LHC energies, the most relevant effect is shadowing: a reduction of the parton densities at low $x$, which intensifies when $Q^2$ decreases and the nucleus mass number $A$ increases.
This effect, induced by the high phase-space density of small-$x$ partons, can be described  within the collinear factorisation framework using phenomenological parametrisations of the 
modification of the PDFs (denoted as nPDFs)~\cite{Eskola:2009uj,Hirai:2007sx,deFlorian:2003qf}. If the parton phase-space reaches saturation, PDF evolution equations are not applicable, and the most appropriate theoretical description is the Colour Glass Condensate effective theory (CGC)~\cite{Gelis:2010nm}. The modification of the small-$x$ parton dynamics can significantly reduce charm-hadron production at low \pt. 
Furthermore, the multiple scattering of partons in the nucleus before and/or after the hard scattering can modify the kinematic distribution of the produced hadrons: partons can lose energy in the initial stages of the collision via initial-state radiation~\cite{Vitev:2007ve}, or experience transverse momentum broadening due to multiple soft collisions before the heavy-quark pair is produced~\cite{Wang:1998ww, Kopeliovich:2002yh}. These initial-state effects are expected to have a small impact on charm-hadron production at high $\pt$ ($\pt>3$--$4$~\GeV/$c$), but they can induce a significant modification of the yield and momentum distribution in the lower \pt region.

The nuclear modification factor \RpPb (the ratio of the cross section in \pPb collisions to that in pp interactions scaled by the mass number of the Pb nucleus) of D mesons measured by ALICE in p--Pb collisions at a centre-of-mass energy per nucleon--nucleon collision \snn = 5.02 \TeV is consistent with unity for $0<\pt<36$~\GeVc~\cite{ALICE:2019fhe}, suggesting that the cold-nuclear-matter effects that influence charm-meson production at midrapidity are small. However, measurements of \Lc baryons in p--Pb collisions~\cite{ALICE:2020wla} indicate a \pt-dependent \RpPb, with values lower than unity for $0<\pt<2~$\GeVc and systematically above unity for $\pt>2$~\GeVc, indicating an
increase in the \Lc mean \pt in \pPb collisions with respect to pp collisions.
A POWHEG+PYTHIA~6 simulation, which is coupled with the EPPS16 nPDF set~\cite{Eskola:2016oht} for p–Pb collisions, is in fair agreement with the \Lc measurements for $\pt<3$~\GeVc, but it does not describe the \RpPb increase above unity in the region $4<\pt<8$~\GeVc.
Measurements in the light-flavour sector in high-multiplicity pp and p--Pb collisions at different energies have revealed strong long-range correlations~\cite{ALICE:2022wpn}, resembling those measured in \PbPb collisions, the latter being understood as due to a collective flow related to the formation of a deconfined QCD medium, the quark--gluon plasma.
A modification of the \pt shape as a function of multiplicity is observed in the strangeness sector by the ALICE and CMS Collaborations in p--Pb collisions~\cite{ALICE:2013wgn, CMS:2019isl} and is consistent with the effect of radial flow in hydrodynamic models such as EPOS LHC~\cite{PhysRevC.92.034906}. In this picture, particles of larger mass are boosted to higher transverse momenta by a common velocity field~\cite{PhysRevC.48.2462}. In the heavy-flavour sector, differential studies of \Lc and \Dzero production as a function of charged-particle multiplicity in pp collisions at \s = 13 \TeV by ALICE~\cite{ALICE:2021npz} revealed a multiplicity dependence of baryon-to-meson yield ratios. In addition, baryon production at intermediate \pt may be enhanced due to hadronisation via quark recombination~\cite{PhysRevLett.90.202303, Beraudo:2023nlq, Minissale:2020bif}. This can be further examined by measuring the \xicz-baryon production in p--Pb collisions and comparing it with the results published in pp collisions~\cite{ALICE:2021psx,ALICE:2021bli}. 

In this paper, the measurement of the \pt-differential production cross section of prompt \xicz baryons in the region 2 $< \pt <$ 12 GeV/$c$ in p--Pb collisions at $\sqrt{s_{\rm NN}}$ = 5.02 \TeV is reported. The term \emph{prompt} refers to charm-hadrons produced directly in the hadronisation of a charm quark or the strong decay of a directly produced excited charm-hadron state, in contrast to \emph{feed-down} charm-hadrons, produced in the decay of a hadron containing a beauty quark. To gauge possible modifications of the \pt distribution of promptly produced \xicz baryons due to the presence of nuclei in the collision system, the nuclear modification factor is also evaluated. Furthermore, information on the hadronisation process can be extracted by comparing the production cross sections of different hadron species. The \pt-dependent yield ratios of \xiczDz and \xicz/\Lc are reported in this article. Finally, by integrating the \pt-differential results and extrapolating them to \pt~=~0, the \pt-integrated prompt \xicz production cross section is computed. This allows the calculation of the charm fragmentation fractions and of the total charm cross section in p--Pb collisions, reported in Ref.~\cite{companionLetter}.

\section{Experimental setup and data samples} \label{sec:datasample}

The ALICE detector system and its performance are described in detail in Refs.~\cite{ALICE:2008ngc,ALICE:2014sbx}. The reconstruction of charm baryons from their hadronic decay products at midrapidity primarily relies on the Inner Tracking System (ITS)~\cite{ALICE:2010tia} for the reconstruction of charged-particle trajectories and determination of primary and decay vertices, the Time Projection Chamber (TPC)~\cite{Alme_2010} for track reconstruction and particle identification (PID) through specific energy loss measurements, and the Time-Of-Flight detector (TOF)~\cite{Akindinov:2013tea}, which extends the PID capabilities of the TPC by measuring the flight time of the charged particles from the interaction point. These detectors are located inside a solenoidal magnet of field strength 0.5~T directed along the beam axis. In addition, the two V0 scintillator arrays~\cite{ALICE:2013axi} are used to trigger collision events and determine the luminosity when used in conjunction with the T0 detector~\cite{ALICE:2016ovj}. The Zero-Degree Calorimeter (ZDC) is employed for offline event selection in \pPb collisions~\cite{ALICE:2014sbx}.

The analysis was performed in the pseudorapidity interval $|\eta|<0.8$ on data from \pPb collisions at \fivenn collected with a minimum-bias (MB) trigger during Run 2 of the LHC. For \pPb collisions, the rapidity in the nucleon--nucleon centre-of-mass system $(y_\mathrm{cms})$ is shifted by 0.46 units in the direction of the proton beam due to the energy asymmetry of the colliding beams. The results are reported for the rapidity interval $|y_\mathrm{lab}|<0.5$ in the laboratory system, which corresponds to $-0.96<y_\mathrm{cms}<0.04$.

The MB trigger requires a pair of coincident signals in the two V0 scintillator arrays, which are located on each side of the interaction point.  Further offline selections were applied to suppress the background originating from beam--gas collisions and other machine-related background sources~\cite{ALICE:2020swj}. In order to maintain uniform ITS acceptance in pseudorapidity, only events with a reconstructed vertex position within 10 cm along the beam axis from the nominal interaction point were analysed. The primary vertex position was determined using tracks reconstructed in the TPC and ITS detectors. Events with multiple interaction vertices reconstructed from TPC and ITS tracks were tagged as pileup from several collisions and removed from the analysed sample~\cite{ALICE:2014sbx}. Using these selection criteria, the p--Pb sample comprised approximately 600 million events, corresponding to an integrated luminosity of $\Lint=287\pm11~\mubinv$.

\section{Data analysis} \label{sec:methods}
In this analysis, \xicz baryons were reconstructed via the decay channel $\xicz \rightarrow {\rm \pi^{+}}\Xi^{-}$ and its charge conjugate, with branching ratio BR $=$ (1.43 $\pm$ 0.32)\%~\cite{pdg2022}. The $\Xi^{-}$ baryons were selected from the decay chain $\Xi^{-} \rightarrow \pi^{-}\Lambda$, BR~$=$~($99.887 \pm 0.035$)\%, followed by $\Lambda \rightarrow {\rm\pi^{-} p}$, BR $=$ ($63.9 \pm 0.5$)\%~\cite{pdg2022}. The $\Xi^{-}$ and $\Lambda$ baryons were reconstructed by exploiting their characteristic decay topologies as reported in Refs.~\cite{Acharya:2019kyh,Acharya:2017lwf}. 
Charged-particle tracks and particle-decay vertices were reconstructed using the ITS and the TPC. The particle trajectories in the vicinity of the primary vertex and the decay vertices were reconstructed with the KFParticle package~\cite{kfparticle}, which allows a direct estimate of their parameters and the associated uncertainties. The package allows one to set constraints on the mass and the production point of the reconstructed particles, using information about the uncertainties of the decay-product trajectories to improve the reconstruction accuracy of the parent particle. The mass constraint
improves the mass and momentum reconstruction of the particle, while the production point constraint
helps to determine whether the particle is coming from a given vertex. These constraints were applied
to the $\Lambda$ and $\Xi^{-}$ decay vertices in the \xicz decay chain reconstruction.  

Several selection criteria were applied for the initial filtering of \xicz candidates.
To ensure good quality of the tracks used to reconstruct the \xicz candidates, track quality and kinematic selection criteria were applied. The tracks were required to be within the pseudorapidity interval $|\eta| < $ 0.8 and to have crossed at least 70 TPC pad rows out of a maximum of 159. The number of clusters in the TPC used for the energy loss determination was required to be larger than 50 to enhance the precision of the mean specific energy loss (\dEdx) measurement. Moreover, the candidate tracks of the $\pi^+$ produced in the \xicz decay were required to have a minimum of three (out of six) hits in the ITS.

The PID selections were based on the difference between the measured and expected detector signals for a given particle species hypothesis in units of the detector resolution ($n^\mathrm{det}_\sigma$). For candidate tracks of the pions from the $\Lambda$ and \X decays and of the proton from the $\Lambda$ decay, a selection on the measured d$E$/d$x$ in the TPC of $|n^\mathrm{TPC}_\sigma|<4$ was applied. For the pion from the \X decay, the transverse momentum was required to be larger than 150~\MeVc, and, if the track has an associated hit in the TOF detector, a further PID selection of $|n^\mathrm{TOF}_\sigma|<5$ was applied to its flight time.
The deviation of the measured invariant masses from the world-average value of the $\Lambda$ and \X masses~\cite{pdg2022} was required to be within 10~\MeVmass. To further reduce the background in the \xicz candidate sample, a selection on the \pt of the pion produced in the \xicz decay was applied. It was required to be greater than 1.6, 1.2, 1.0, and 1.0 \GeVc in the 2--4, 4--6, 6--8, and 8--12 \xicz \pt intervals, respectively.

After applying the selections described above, further rejection of
background \xicz candidates was obtained using a boosted decision tree (BDT) algorithm. The BDT implementation provided by the XGBoost library was used~\cite{chen2016xgboost,hipe4ml}. During the training process, the BDT is presented with labelled samples, where the true classifications of signal and background are known. The algorithm is trained to optimise the classification of the two classes based on the differences in the topology, kinematics, and PID information. Once the BDT is trained, it is applied to the data sample where the classifications are unknown.
With the machine learning approach, multiple selection criteria are combined into a single response variable representing the algorithm's confidence in classifying a candidate as a true \xicz baryon. After applying a trained BDT model to the data sample, a selection in the BDT response was applied to reduce the large combinatorial background. Independent BDTs were trained for each \pt interval in the
analysis. The signal samples for the training were obtained from simulated events using the PYTHIA~8.243~\cite{Sjostrand:2014zea} Monte Carlo (MC) generator with the Monash tune~\cite{Skands:2014pea} embedded into an underlying \pPb collision generated with HIJING 1.36~\cite{wang1991hijing}. The transport of simulated particles within the detector was performed with the GEANT 3 package~\cite{Brun:1082634}. The LHC beam conditions and the conditions of the ALICE detectors in terms of active channels, gain, noise level, and alignment, and their evolution with time during the data taking, were taken into account in the simulations. Each PYTHIA~8 event was required to contain a c$\overline{\mathrm{c}}$ quark pair, with at least one hadronising into a \xicz baryon, which is forced to decay via the decay channel of interest. Only prompt \xicz signal candidates were selected for the training, while feed-down ones were not used since they have a different decay vertex topology. The background sample was selected from a fraction of real data using the same selection criteria described above, with the additional requirement that the invariant mass of the \xicz candidate was within the intervals $2.17 < M < 2.39$~\GeVmass or $ 2.55 < M < 2.77$~\GeVmass (sidebands) to ensure that the signal region was excluded.

The training variables in the BDT related to the \X candidates were i) the reconstructed invariant mass, ii) the pointing angle of its momentum to the primary vertex, i.e.\ the angle between the momentum vector of the reconstructed \X particle and the vector pointing from the reconstructed primary vertex to the \X decay vertex,  iii) the decay length normalised by its uncertainty, iv) a normalised $\chi^2$ value, which is obtained by evaluating whether the momentum vector of the \X candidate points back to the reconstructed primary vertex, and is provided by the KFParticle package. The training variables describing the pion emerging from the \xicz decay were the $n\sigma^{\rm TOF}_{\pi} $, the $n\sigma^{\rm TPC}_{\pi}$ and the distance of closest approach to the primary vertex. The selection on the BDT response was tuned in each \pt interval to maximise the expected statistical significance, which is calculated using i) an estimated value for the signal extrapolated from the \xicz production cross section reported in Ref.~\cite{ALICE:2021psx} using a \LT fit, multiplied by the reconstruction and selection efficiencies for each BDT selection threshold,  ii) an estimate of the background within the signal region obtained by interpolating a fit to the invariant mass distribution in the sidebands of the \xicz signal region.

\begin{figure} [htb]
    \centering
    \includegraphics[width=.49\linewidth]{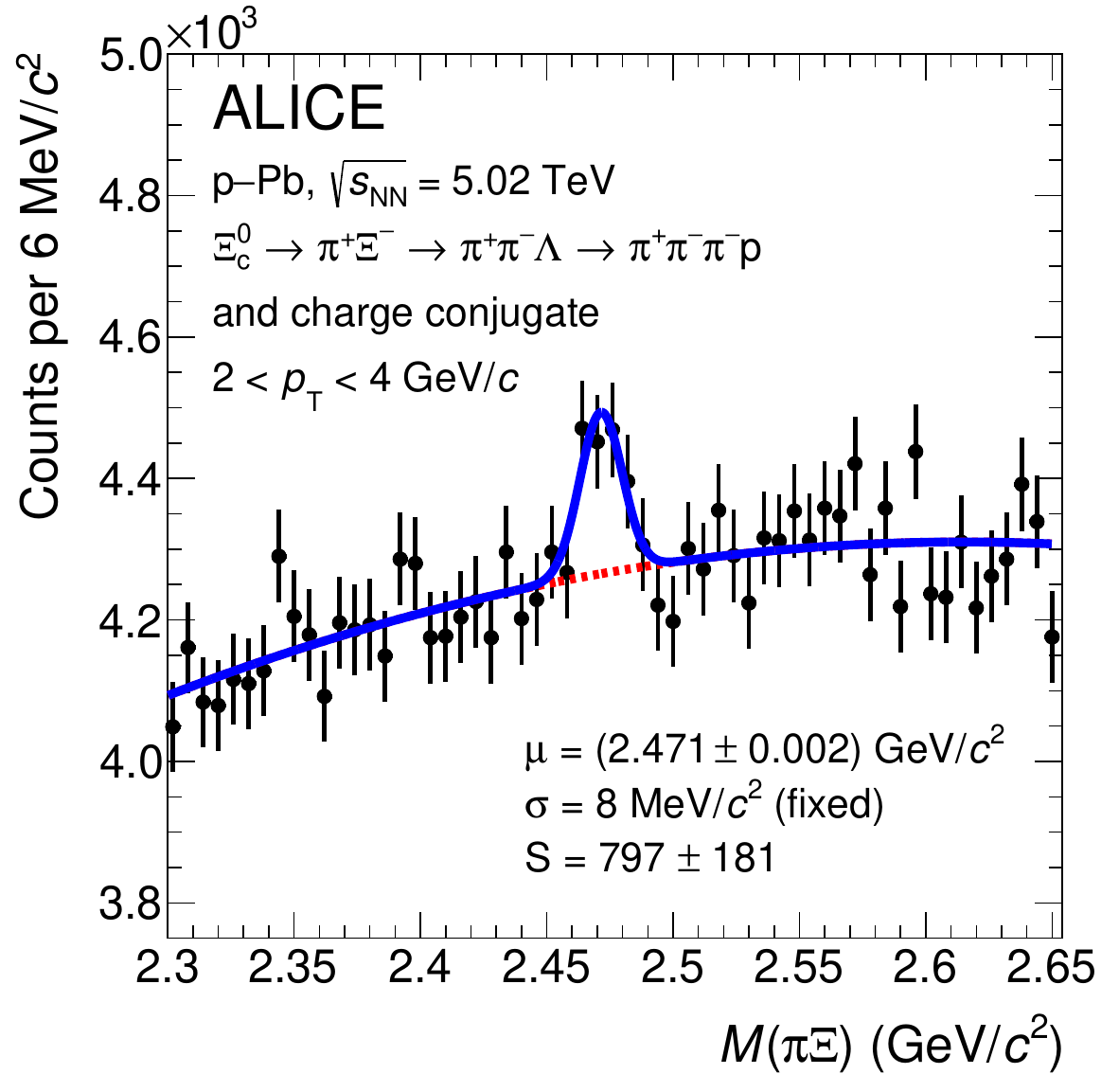}
    \includegraphics[width=.49\linewidth]{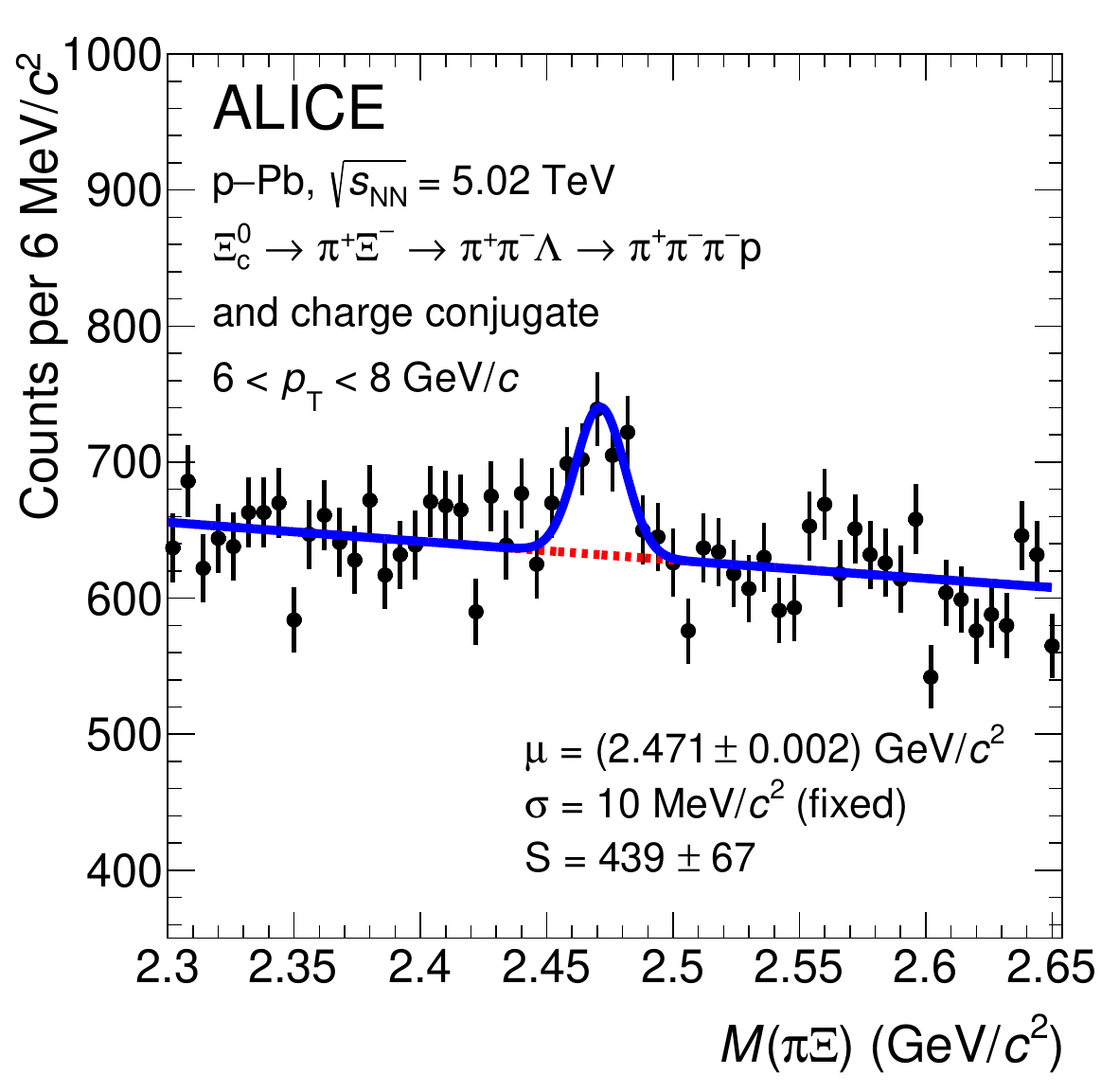}
    \caption{Invariant mass distributions of $\xicz \rightarrow {\rm \pi^{+}}\Xi^{-}$ candidates (and charge conjugates) in $2 < \pt < 4$~ \GeVc (left), and in $6 < \pt < 8$~\GeVc (right) in \pPb collisions at \fivenn. The red dashed curves represent the background fit functions, and the blue curves represent the total fit functions.}
    \label{fig:InvMass}
\end{figure}

After applying the BDT selections, the raw \xicz yield in each \pt interval under study was obtained by fitting the invariant mass distribution of the candidates. Examples of those distributions are shown in Fig.~\ref{fig:InvMass} for the \pt intervals 2--4 and 6--8 GeV/$c$.
A Gaussian function was used to model the signal peak and an exponential (for \pt $> 4$ \GeVc) or parabolic function (in the $2 < \pt <4$ \GeVc interval) was used to model the background. Due to the small signal-to-background ratio, the standard deviation of the Gaussian signal function was fixed to the value obtained from simulations to improve the fit stability. A \xicz signal, with a statistical significance larger than 3, could be extracted in the four considered \pt intervals in the range $2<\pt<12$~\GeVc.

\section{Corrections} \label{sec:corrections}
\begin{figure}[htb]
    \centering
    \includegraphics[width=0.5\linewidth]{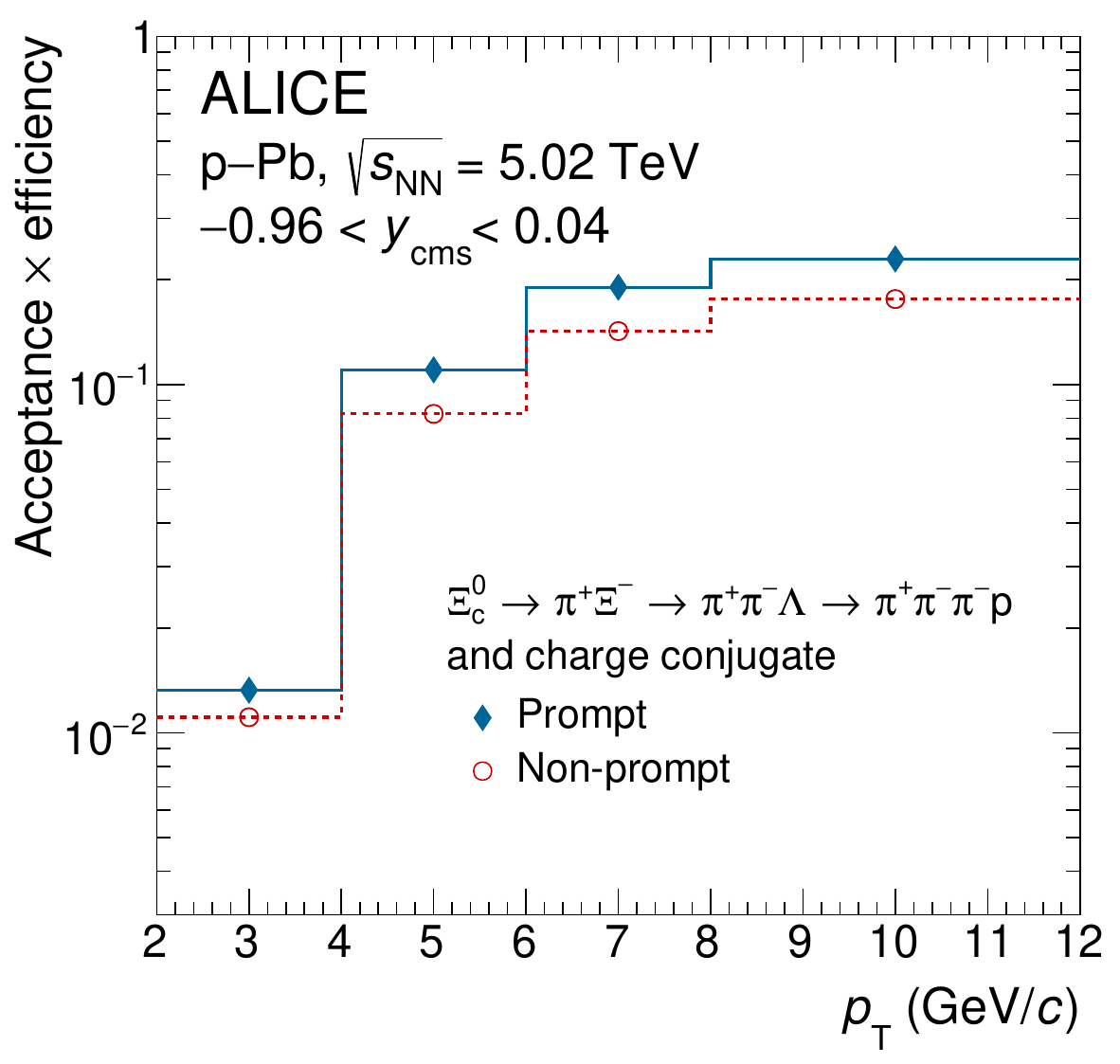}
    \caption{Product of detector acceptance and efficiency for \xicz baryons in \pPb collisions at \fivenn as a function of \pt. The solid line corresponds to the $(\mathrm{Acc}\times\epsilon)$ for prompt \xicz, while the dotted line represents $(\mathrm{Acc}\times\epsilon)$ for \xicz baryons originating from beauty-hadron decays. The statistical uncertainties are smaller than the marker size.}
    \label{fig:Eff}
\end{figure}

The \pt-differential production cross section of prompt \xicz baryons per unit of rapidity in the interval $-0.96 < y_{\mathrm{cms}} < 0.04$ was calculated from the raw yields as

\begin{equation}
  \frac{\mathrm{d^2}\sigma}{\mathrm{d}y~\mathrm{d}p_{\mathrm{T}}} =
  \frac{1}{2}
  \frac{N^{\mathrm{\xicz+\overline{\xicz}}}_{\mathrm{raw}}(p_{\mathrm{T}})\times f_{\mathrm{prompt}}(p_{\mathrm{T}})}
  {\Delta y_\mathrm{lab}\,\Delta p_{\mathrm{T}} \times(\mathrm{Acc}\times\epsilon)_{\mathrm{prompt}}(p_{\mathrm{T}})\times\mathrm{BR}\times \Lint },
  \label{eq:xsec}
 \end{equation}

    where $N^{\mathrm{\xicz+\overline{\xicz}}}_{\mathrm{raw}}$ is the raw yield (sum of particles and antiparticles) extracted from the invariant mass fit, $f_{\mathrm{prompt}}$ is the fraction of prompt \xicz in the measured raw yield, BR is the branching ratio of the considered decay chain, and \Lint is the integrated luminosity. The factor 2 accounts for the presence of particles and antiparticles in the raw yields, and $\Delta y_\mathrm{lab}\,\Delta p_{\mathrm{T}}$ accounts for the widths of the rapidity and transverse momentum intervals. The measurement of \xicz is performed for $\Delta y_\mathrm{lab} = 1.6$, under the assumption that the cross section per unit of rapidity of \xicz baryons does not significantly change between $\lvert y_\mathrm{lab} \rvert < 0.5$ and $\lvert y_\mathrm{lab} \rvert < 0.8$. This has been verified using PYTHIA 8 simulations~\cite{Sjostrand:2014zea} and FONLL pQCD calculations~\cite{Cacciari:1998it, fonllcalc1}. The factor $(\mathrm{Acc}\times\epsilon)_{\mathrm{prompt}}$ is the product of the geometrical acceptance (Acc) and the reconstruction and selection efficiency ($\epsilon$) for prompt \xicz candidates and is evaluated using the PYTHIA 8 simulations described in Sec.~\ref{sec:methods}. To account for differences in the \pt-distribution of prompt \xicz baryons between the MC simulation and the actual data distribution, weighting factors were applied to the generated \xicz \pt-distribution before calculating $(\mathrm{Acc}\times\epsilon)_{\mathrm{prompt}}$. These weighting factors were determined by comparing the \pt-distribution of \xicz generated from the Quark Combination Model (QCM)~\cite{Li:2017zuj} to the PYTHIA~8 distribution described above. The QCM was chosen because it best reproduces the measured \xicz \pt distribution.
Figure~\ref{fig:Eff} shows the final $(\mathrm{Acc}\times\epsilon)$ correction factors of prompt and feed-down \xicz as a function of \pt. The $(\mathrm{Acc}\times\epsilon)$ of prompt \xicz baryons is slightly larger than that of feed-down baryons. This difference arises from the inclusion of the distance of closest approach (DCA) to the primary vertex of the pion produced in the \xicz decay as one of the variables in the BDT training. Since the pions from decays of feed-down \xicz present a wider DCA distribution as compared to those from prompt \xicz, due to the displacement by a few hundred micrometres of the feed-down \xicz decay vertices, prompt signals tend to exhibit higher BDT output scores compared to feed-down signals.

The fraction of prompt baryons in the raw \xicz yield extracted from the selected candidate sample was calculated as:
\begin{equation}
\begin{split}
    f_{\mathrm{prompt}} &= 1 - \frac{N^{\mathrm{\xicz+\overline{\xicz}}}_{\mathrm{feed\text{-}down}}}{N^{\mathrm{\xicz+\overline{\xicz}}}_{\mathrm{raw}}} = \\
    & = 
    1 - \frac{1}{N^{\mathrm{\xicz+\overline{\xicz}}}_{\mathrm{raw}}} \frac{\mathrm{d^2}\sigma^{\xicz}_{\mathrm{feed\text{-}down}}}{\mathrm{d}y~\mathrm{d}p_{\mathrm{T}}}\times 2 \times (\mathrm{Acc}\times\epsilon)_{\mathrm{feed\text{-}down}} \times R_{\mathrm{pPb,~feed\text{-}down}} \times \Delta p_{\mathrm{T}} \times \Delta y_{\rm lab} \times \mathrm{BR} \times \Lint  
\end{split}
\end{equation}

The yield of feed-down \xicz ($\mathrm{d^2}\sigma^{\xicz}_{\mathrm{feed\text{-}down}}/\mathrm{d}\pt \mathrm{d} y$) is estimated starting from the cross section of \lambdac baryons originating from \lambdab decays, which is obtained using the beauty-quark production cross section from FONLL calculations~\cite{fonllcalc1,fonllcalc2}, the fraction of beauty quarks fragmenting into beauty hadrons taken from the LHCb measurement of beauty fragmentation fractions in pp collisions at \thirteen~\cite{LHCb:2019fns}, and the decay kinematics of beauty hadrons decaying into a final state with a \lambdac, which is taken from PYTHIA 8 ($(\mathrm{d^2}\sigma/\mathrm{d}\pt \mathrm{d} y)^{\lambdac}_{\mathrm{feed\text{-}down,FONLL}}$). The predicted feed-down \lambdac cross section is scaled by the ratio of the measured \pt-differential production cross sections of prompt \xicz~\cite{ALICE:2021psx} $(\mathrm{d^2}\sigma/\mathrm{d}\pt \mathrm{d} y)^{\xicz}_{\mathrm{prompt}}$,
and prompt \lambdac~\cite{ALICE:2020wfu},  where the $(\mathrm{d^2}\sigma/\mathrm{d}\pt \mathrm{d} y)^{\lambdac}_{\mathrm{prompt}}$ was measured in pp collisions at \five to obtain an estimation for the \pt-differential feed-down \xicz production cross section

\begin{equation}
    \frac{\mathrm{d^2}\sigma^{\xicz}_{\mathrm{feed\text{-}down}}}{\mathrm{d}y~\mathrm{d}p_{\mathrm{T}}} = \left(\frac{\mathrm{d^2} \sigma}{\mathrm{d}\pt \mathrm{d} y}\right)^{\lambdac}_{\mathrm{feed\text{-}down,FONLL}} \times \frac{(\mathrm{d^2}\sigma/\mathrm{d}\pt \mathrm{d} y)^{\xicz}_{\mathrm{prompt}}}{(\mathrm{d^2}\sigma/\mathrm{d}\pt \mathrm{d} y)^{\lambdac}_{\mathrm{prompt}}} 
\end{equation}

The scaling factor $(\mathrm{d^2}\sigma/\mathrm{d}\pt \mathrm{d} y)^{\xicz}_{\mathrm{prompt}}/(\mathrm{d^2}\sigma/\mathrm{d}\pt \mathrm{d} y)^{\lambdac}_{\mathrm{prompt}}$ approximates the term

 \begin{equation}
\frac{\sum_{\mathrm{H_B}} \mathrm{b} \rightarrow \mathrm{H_{b}} \rightarrow \mathrm{\Xi_{c}}}{\sum_{\mathrm{H_B}} \mathrm{b} \rightarrow \mathrm{H_{b}} \rightarrow \mathrm{\Lambda_{c}}} \simeq
\frac{\mathrm{b} \rightarrow \mathrm{\Xi_{b}} \rightarrow \mathrm{\Xi_{c}}}{\mathrm{b} \rightarrow \mathrm{\Lambda_{b}}\rightarrow \mathrm{\Lambda_{c}}}
 \end{equation}
 
considering that the only beauty baryons that contribute to the yield of $\Xi_{\rm c}$ ($\lambdac$) are $\Xi_{\rm b}$ ($\lambdab)$.
This scaling relies on the assumptions that the \pt shapes of the feed-down \lambdac and \xicz production cross sections are similar, and the ratio of production cross sections of \xicz and \lambdac is similar for prompt and feed-down baryons.
Finally, a hypothesis on the value of the nuclear modification factor $R_{\mathrm{pPb}}$ of feed-down \xicz is needed. It is assumed that the $R_{\mathrm{pPb}}$ of prompt \xicz is equal to that of prompt \lambdac and that the $R_{\mathrm{pPb}}$ of prompt and feed-down \xicz are equal. The evaluated prompt fraction ranges between 0.91 and 0.96 depending on the \pt interval.

\section{Systematic uncertainties} \label{sec:systematics}
The contributions to the systematic uncertainty on the prompt \xicz production cross section are summarised in Table~\ref{tab:syst_uncert}.

\begin{table}[htp]
	\begin{center}
            \vspace{0.2cm}
 		\caption{Systematic uncertainties on the prompt \xicz production cross section. }
		\label{tab:syst_uncert}
		\begin{tabular}{ l c  c c c }
    \toprule
    \toprule
   \multicolumn{1}{c|}{}& \multicolumn{4}{c}{$\pt$ interval (GeV/$c$)} \\
    
	\multicolumn{1}{c|}{}   & [2,4] & [4,6] & [6,8]& [8,12]\\
        \midrule
        Raw yield extraction    &   14\%&    7\%&   8\% &   6\% \\
        Tracking efficiency     &    4\%&    2\%&   3\% &   3\% \\
        Matching efficiency     &    1\%&    1\%&   1\% &   1\% \\
        BDT selection efficiency&    6\%&   10\%&   5\% &   3\% \\
        
        MC \pt shape            &    5\%&    1\%&   1\% &   1\% \\
        Feed-down subtraction   &   $_{-6}^{+5}$\%    & $^{+8}_{-11}$\%   &     $^{+10}_{-15}$\%        & $^{+6}_{-9}$\%      \\
        \midrule
        Branching ratio &  \multicolumn{4}{c}{22\%}\\
        Luminosity  & \multicolumn{4}{c}{3.7\%}\\
        \bottomrule
        \bottomrule
		\end{tabular}
	\end{center}
\end{table}

The systematic uncertainty on the raw yield extraction was evaluated by repeating the fit to the invariant mass distributions varying: i) the function used to describe the background, ii) the minimum and maximum of the invariant mass intervals considered for the fit, and iii) the Gaussian width of the mass peak by $\pm10$\% compared to the value obtained from simulations. The systematic uncertainty was assigned by adding in quadrature the root mean square error (RMS) of the resulting distribution of the raw yield values and the shift of the mean of the distribution with respect to the value obtained with the default fit configuration. 

The systematic uncertainty on the track reconstruction efficiency was estimated by i) varying the track selection criteria in the TPC and ii) comparing the ITS-TPC matching efficiency in data and simulations. These contributions are reported as “Tracking efficiency” and “Matching effinciency” in Table~\ref{tab:syst_uncert}, respectively.
The first contribution to the tracking systematic uncertainty was defined as the RMS of the \xicz cross section values obtained by repeating the analysis with different TPC track selection criteria.
The matching efficiency affects only the pion from the \xicz decay since no ITS condition was imposed for tracks coming from the \X and $\Lambda$ decays. The per-track uncertainty on the matching efficiency is \pt dependent, and it was propagated to the \xicz taking into account the decay kinematics.  

The systematic uncertainty on the \xicz selection efficiency arises due to possible differences between the real detector resolutions and alignment, and their description in the simulation. This uncertainty was assessed by comparing the production cross sections obtained using different selection criteria. In particular, the selections on the BDT outputs were varied in a range corresponding to a modification of about 40\% in the efficiency for the prompt \xicz. The systematic uncertainty was assigned as the RMS of the resulting production cross section distribution. An additional source of systematic uncertainty was assigned due to the dependence of the efficiencies on the generated \pt distribution of \xicz in the simulation (“MC \pt shape” in Table~\ref{tab:syst_uncert}). To estimate this effect, the efficiencies were evaluated modifying the weights to match the \pt spectrum obtained from PYTHIA~8 simulations using a tune with colour-reconnection topologies beyond the leading-colour approximation~\cite{Christiansen:2015yqa}, which includes so-called “junctions” that fragment into baryons and
lead to an increased baryon production with respect to the Monash tune. An uncertainty was assigned in each \pt interval based on the difference between the central and the varied efficiency. 

The systematic uncertainty on the subtraction of feed-down from beauty-hadron decays was estimated by varying: i) the factorisation and renormalisation scales, and the beauty quark mass in the FONLL calculations (according to the prescriptions in Ref.~\cite{Cacciari:2005rk}), and the hypothesis on the fragmentation functions within the uncertainties of the LHCb measurement (taken from Refs.~\cite{Gladilin:2014tba}), ii) the $\xicz$/$\Lambda^+_{\rm c}$ ratio, scaling it up by a factor of 2 and down by a factor tuned to accommodate the $\Xi_{\rm b}^{\rm -}/\Lambda_{\rm b}^{\rm 0}$ ratio measured by the LHCb Collaboration~\cite{LHCb:2019fns}, iii) the hypothesis on the $R_{\mathrm{pPb}}$ of feed-down \xicz in the range $0.9 < R_{\mathrm{pPb, feed-down}}/R_{\mathrm{pPb, prompt}} < 2$. This range is chosen to cover the uncertainties of the measured \lambdac $R_{\mathrm{pPb}}$. The theoretical calculation from QCM was also considered for the systematic variation of the $R_{\mathrm{pPb}}$. 

The uncertainty on the luminosity measurement is 3.7\% for \pPb collisions~\cite{ALICE:2014gvw} and the uncertainty on the decay chain branching ratio is 22\%~\cite{pdg2022}.

\section{Results} \label{sec:results}
\begin{figure} [!tb]
    \centering
    \includegraphics[width=0.49\linewidth]{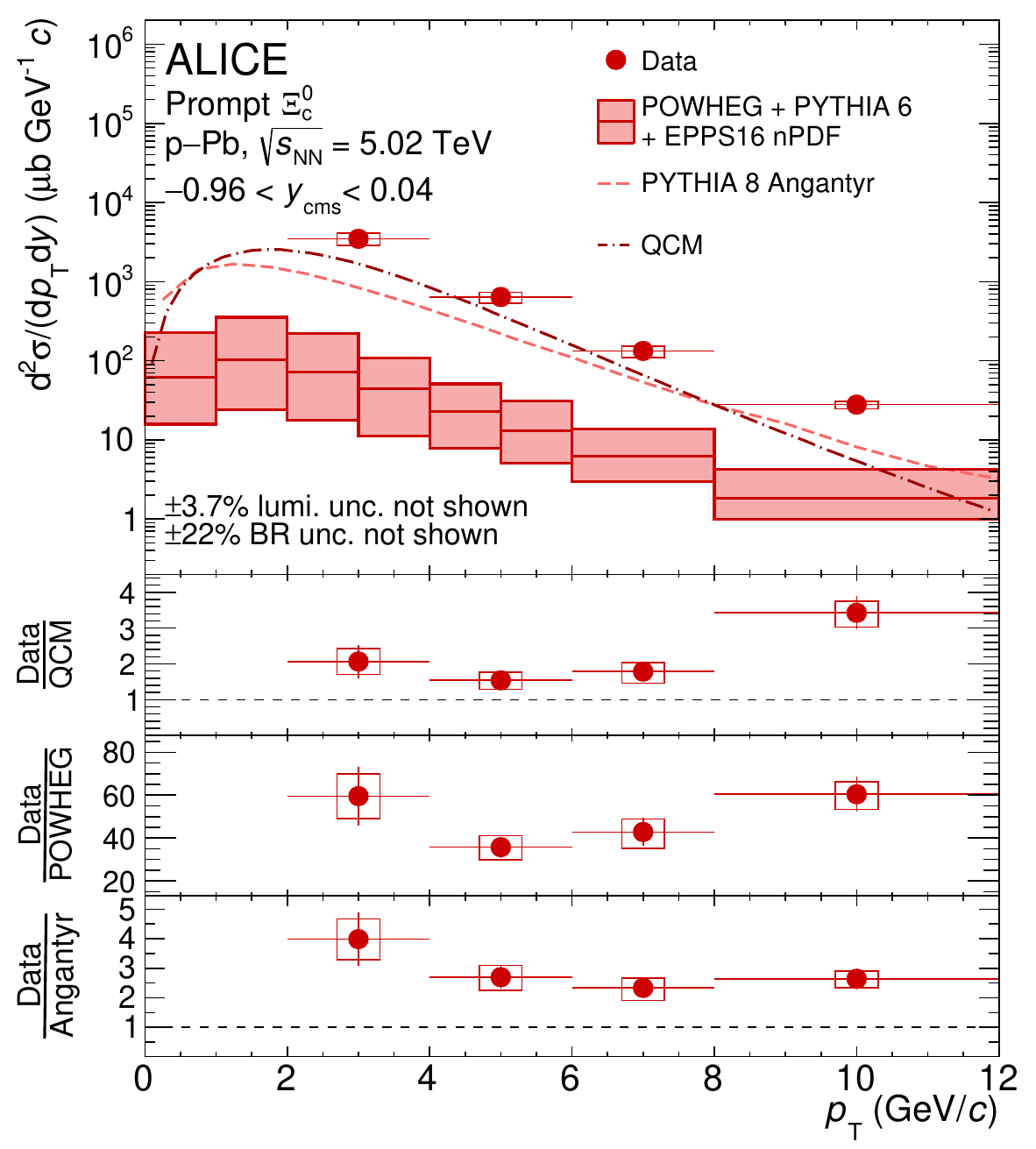}
    \includegraphics[width=0.49\linewidth]{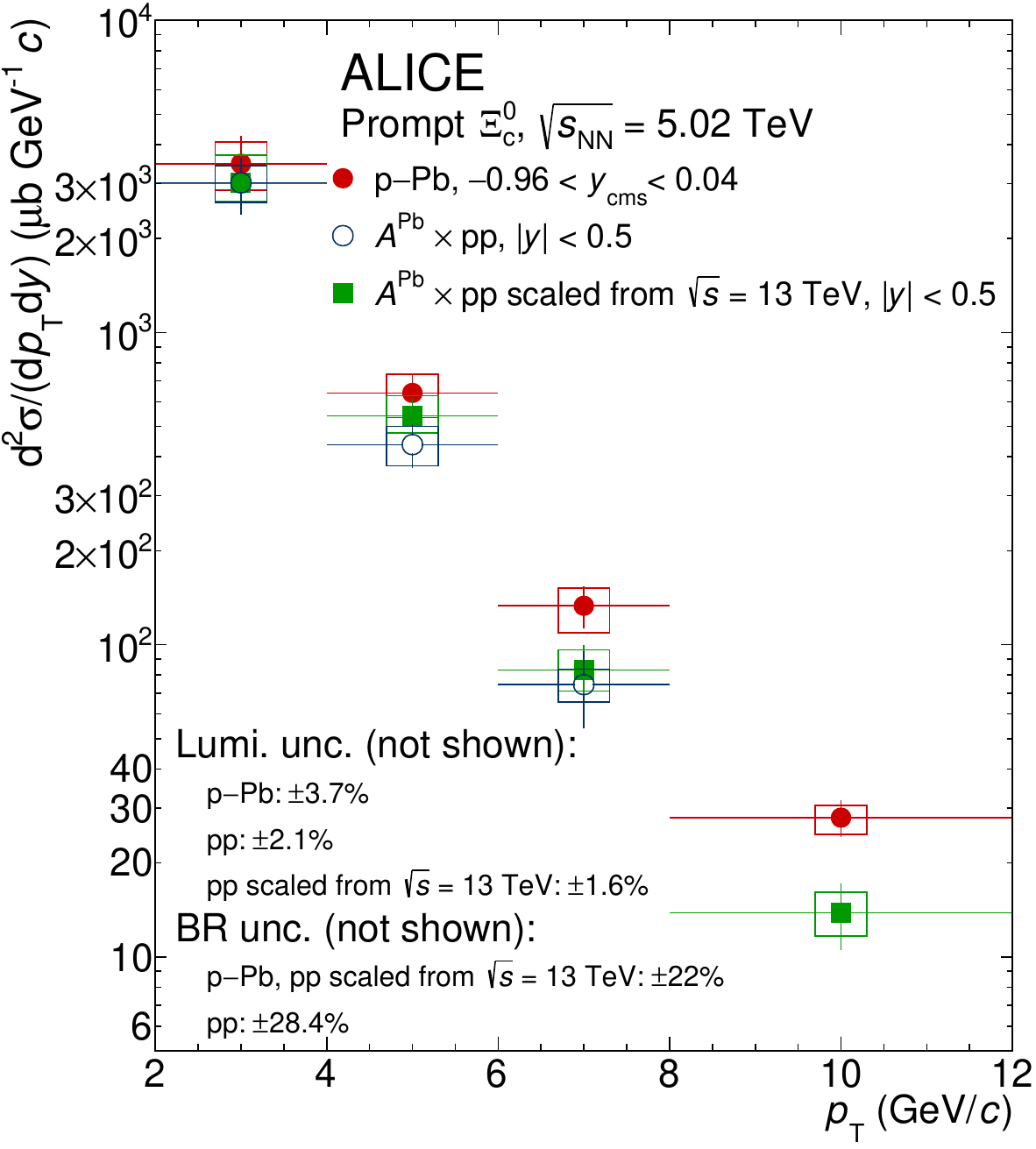}
    \caption{Prompt \xicz-baryon \pt-differential cross section in \pPb collisions at \fivenn. The statistical uncertainties are shown as vertical bars and the systematic uncertainties are shown as boxes. Left: comparison to predictions from POWHEG+PYTHIA~6~\cite{Frixione:2007nw, sjostrand2006pythia} simulations matched with the EPPS16 nPDF parameterisation~\cite{Eskola:2016oht}, PYTHIA~8 Angantyr calculations~\cite{Bierlich:2023okq}, and the QCM~\cite{Li:2017zuj}. The uncertainties on the POWHEG calculation are due to the choice of the pQCD scales and the charm quark mass as described in the text. Right: Comparison with the cross section measured in the semileptonic decay channel in \pp collisions at \five~\cite{ALICE:2021psx} and in the hadronic decay channel in \pp collisions at \thirteen~\cite{ALICE:2021bli} scaled to \five (the energy-scaling factor is described in the text). Both the \pp measurements are scaled by the atomic mass number A of the Pb nucleus.}
    \label{fig:CrossSection}
\end{figure}

\begin{sloppypar}
The \pt-differential cross section of prompt \xicz-baryon production in \pPb collisions at \fivenn measured in the transverse momentum interval $2<\pt< 12$~\GeVc and in the rapidity interval ${-0.96 < y_\mathrm{cms} < 0.04}$ is shown in Fig.~\Ref{fig:CrossSection}. In the left panel of Fig.~\Ref{fig:CrossSection}, the measured cross section is compared with POWHEG+PYTHIA~6 calculations and with QCM predictions~\cite{Li:2017zuj}. 
\end{sloppypar}

In POWHEG+PYTHIA~6 calculations, the CT14NLO parton distribution functions~\cite{Dulat:2015mca} are used for the proton, while for the Pb nucleus the nuclear modification of the PDFs is modelled with the EPPS16 nPDF parameterisation~\cite{Eskola:2016oht}.
The factorisation and renormalisation scales, $\mu_{\mathrm{F}}$ and $\mu_{\mathrm{R}}$, were taken to be equal to the transverse mass of the quark, $\mu_{\mathrm{0}}=\sqrt{m^2+p_{\mathrm{T}}^2}$, and the charm-quark mass was set to ${m_{\mathrm{c}}=1.5}$~\GeVmass. The theoretical uncertainties were estimated by varying these scales in the intervals $0.5 \mu_{0} < \mu_{\mathrm{F,R}} < 2 \mu_{\mathrm{0}}$, with $0.5 \mu_{0} < \mu_{\mathrm{R}}/\mu_{\mathrm{F}} < 2 \mu_{\mathrm{0}}$ based on the prescriptions of Ref.~\cite{fonllcalc1}. The uncertainties on the nPDF were not included in the calculation as they are considerably smaller than the scale uncertainties. The POWHEG+PYTHIA~6 calculations underestimate the measurement by about a factor of 50. This discrepancy is mostly attributed to the description of the hadronisation process in PYTHIA~6, which is tuned on $\mathrm{e^+ e^-}$ collisions and underestimates baryon production. This is also supported by the $\xicz/\DZero$ and $\xicz/\Lc$ ratios described in the following. Notably, a large difference between the measured \xicz production and the predictions from PYTHIA~8 with the Monash tune, in which the hadronisation is tuned on $\mathrm{e^+ e^-}$ collisions, was already observed in pp collisions both for the \xicz cross section~\cite{ALICE:2021psx} and the baryon-over-meson ratios~\cite{ALICE:2021psx, ALICE:2021bli}. In the PYTHIA~8 Angantyr calculations~\cite{Bierlich:2023okq} the \pPb collision is treated as a superposition of multiple nucleon--nucleon collisions. In addition, this model implements the colour-reconnection mechanism beyond leading-colour approximation which includes the formation of junctions that fragment into baryons. The junction formation and fragmentation for the colour dipoles containing heavy quarks has been improved. The rope hadronization model, needed to increase the strange quark production in strings breakups, is not yet implemented and its effect is emulated by increasing the overall relative probability of having strange quarks in string breakups. The PYTHIA~8 Angantyr calculations underestimate the measured \xicz production cross section in \pPb collisions by a factor of about 3. The QCM calculations implement charm-quark hadronisation via  a simplified approach to hadronisation by coalescence, developed solely in the momentum space, without considering the coordinate space. The model estimates the \pt distributions of quarks from a fit to the measured $\pi$, K, and D meson spectra, and assumes that hadronisation occurs exclusively through coalescence at all momenta~\cite{Li:2017zuj, Song:2018tpv}. The charm quark is combined with a co-moving light antiquark or two co-moving quarks to form a charm meson or baryon. A free parameter, $R^{\mathrm{(c)}}_{\mathrm{B/M}}=0.425$, characterises the relative production of single-charm baryons to single-charm mesons and it is tuned to reproduce the \lambdac/\DZero ratio measured by ALICE in \pp collisions at \seven ~\cite{ALICE:2017thy}. The relative abundances of the different charm-baryon species are determined by thermal weights from the statistical hadronisation approach~\cite{Andronic:2009sv}. The QCM prediction, tuned on the \lambdac production in pp collisions, underestimates the measured \xicz production cross section in \pPb collisions by a factor of about 2. A similar discrepancy was observed in pp collisions~\cite{ALICE:2021bli} when comparing the results to a statistical hadronisation approach~\cite{He:2019tik}, which also employs thermal weights to determine the abundances of different charm-baryon species~\cite{ALICE:2021bli}.

In the right panel of Fig.~\Ref{fig:CrossSection} the prompt \xicz-baryon production cross section measured in \pPb collisions is compared with the measurement in the semileptonic decay channel in \pp collisions at \five~\cite{ALICE:2021psx} and with the measurement in the hadronic decay channel $\xicz \rightarrow {\rm \pi^{+}}\Xi^{-}$ in \pp collisions at \thirteen~\cite{ALICE:2021bli} scaled to \five. Both the \pp cross sections are scaled by the atomic mass number $A=208$ of the lead nucleus. The prompt \xicz production cross section measured in \pp collisions at \thirteen was first rebinned to match the \pt intervals of the \pPb measurement. In the rebinning, the raw yield uncertainty was considered uncorrelated across \pt intervals, while all other systematic uncertainties were treated as correlated. The \pt-dependent energy-scaling factor from \thirteen to \five was computed as the ratio of \DZero-meson production cross section from FONLL calculations in \pp collisions at the same collision energies~\cite{Averbeck:2011ga}. The ALICE Collaboration has shown that the energy scaling factors for mesons and baryons are compatible within the uncertainties~\cite{ALICE:2023sgl}. 
The uncertainty on the energy-scaling factor was calculated by varying the renormalisation and factorisation scales, the mass of the charm quark, and the nuclear PDFs consistently at the two collision energies. The total uncertainty is calculated as the envelope of the variations. The two results for the \xicz cross sections in \pp collisions at \five are in agreement with each other within uncertainties. The cross section measured via the hadronic decay channel in pp collisions at \thirteen scaled to \five is used as the \pp reference for the calculation of the nuclear modification factor because the two measurements are performed in the same decay channel and the systematic uncertainty due to the branching ratio, which is the dominant systematic uncertainty, cancels out in the ratio. Furthermore, the cross section in \pp collisions at \thirteen was measured in the \pt interval 8--12 GeV$/c$, allowing the computation of the \RpPb up to higher \pt as compared to the measurement in \pp collisions at \five, which extends only up to \pt = 8 GeV$/c$.

\begin{figure} [tb!]
    \centering
    \includegraphics[width=0.5\linewidth]{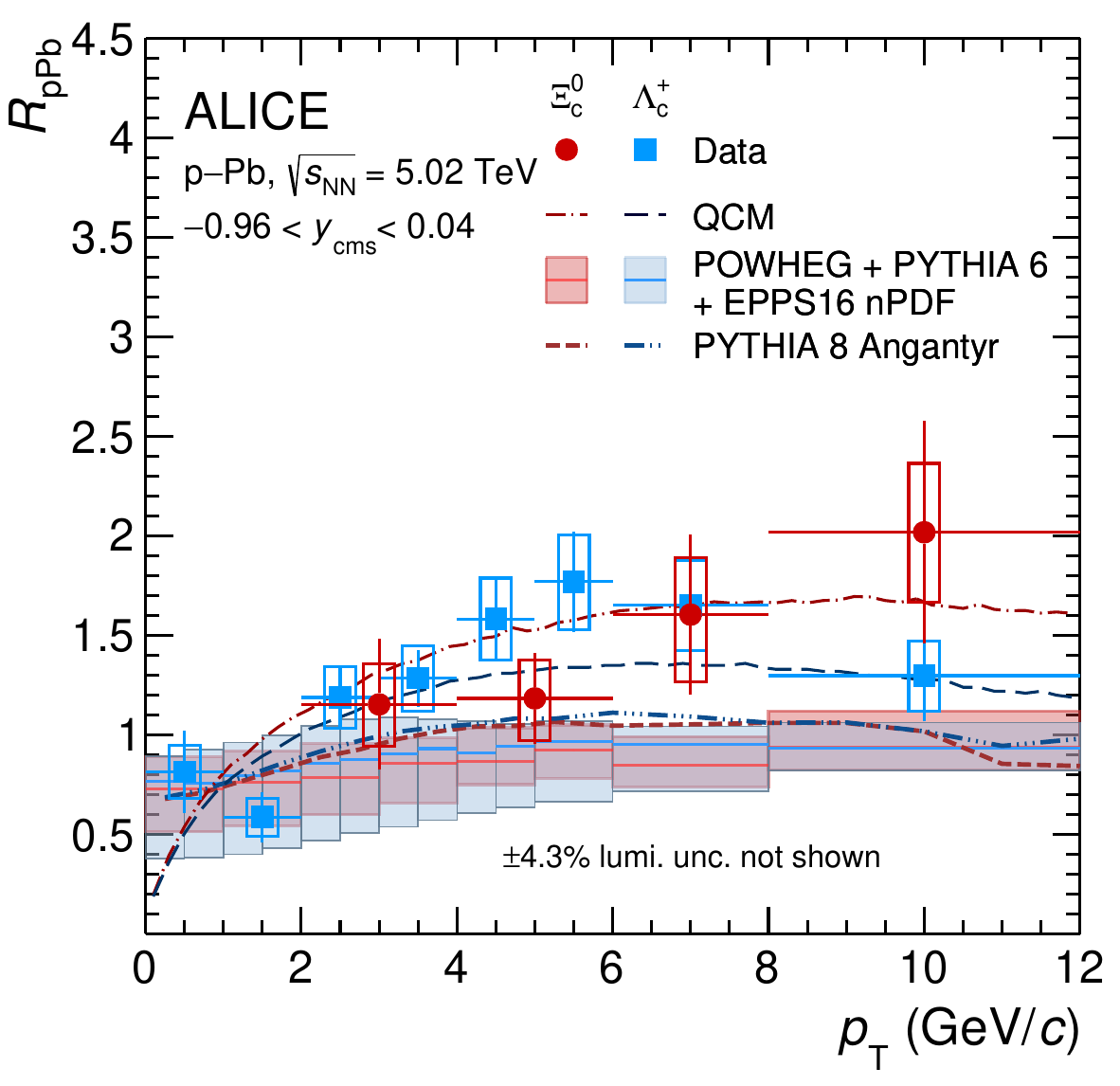}
    \caption{Nuclear modification factor \RpPb of prompt \xicz baryons in \pPb collisions at \fivenn as a function of \pt compared to the \RpPb of \lambdac baryons~\cite{ALICE:2022exq}. The measured \RpPb is also compared to POWHEG+PYTHIA~6 with EPPS16 simulations, to PYTHIA~8 Angantyr calculations~\cite{Bierlich:2023okq}, and to QCM predictions. The statistical uncertainties are shown as vertical bars and the systematic uncertainties are shown as boxes. The uncertainties on the POWHEG calculation are due to the choice of the pQCD scales as described in the text. }
    \label{fig:RpPb}
\end{figure}

To better investigate any effect due to the collision system, the nuclear modification factor \RpPb was calculated as the ratio between the \pt-differential \xicz cross section in \pPb collisions and the reference pp measurement scaled by the nuclear mass number $A $ of the lead nucleus and corrected to account for the rapidity shift between pp and p–Pb collisions using FONLL calculations~\cite{fonllcalc1}. The result is shown in Fig.~\Ref{fig:RpPb}. The systematic uncertainties on the branching ratio and beauty feed-down are treated as fully correlated between the two collision systems, and all other systematic uncertainties are considered as uncorrelated. The central values of the \xicz-baryon \RpPb are larger than unity in the full \pt interval of the measurement, even though all data points are compatible with unity within the large uncertainties (the maximal deviation is 1.5$\sigma$ for the \pt interval 8--12 \GeVc). This also prevents to establish if an increasing trend with \pt is present. The \xicz \RpPb is compatible with the  \lambdac \RpPb~\cite{ALICE:2022exq} within uncertainties pointing to a similar modification of the production of these two baryon species in \pPb collisions with respect to \pp collisions. The measured \RpPb is compared to POWHEG+PYTHIA~6 simulations, PYTHIA~8 Angantyr calculations, and QCM predictions. The only nuclear effect included in the POWHEG + PYTHIA~6 calculations is due to the EPPS16 modification of the PDFs. The resulting \RpPb is lower than unity with a mild \pt dependence, and it reproduces within the large uncertainties the measured \xicz \RpPb. On the other hand, some tension is visible for the \lambdac in the $4 < \pt < 8$~\GeVc interval. The QCM reproduces both measurements within their uncertainties.
The comparison with POWHEG+PYTHIA~6 calculations suggests that the deviation of \RpPb from unity (significant for the \Lc) may have influences beyond the alteration of the PDFs of nucleons within the nuclei's structure compared to those of protons. Additional effects, possibly related to the hadronisation process and the presence of an expanding medium, may play a role, as suggested by the modification of the \Lc \pt shape between pp and \pPb collisions~\cite{ALICE:2022exq}. However, the large uncertainties associated with the \xicz results prevent conclusive interpretations, because the measured \RpPb exhibits a trend compatible with both the \Lc and a flat trend. The \RpPb obtained from PYTHIA~8 Angantyr calculations shows a mild increasing trend with \pt, reaching unity at $\pt\sim 3$~\GeVc for both \xicz and \lambdac and is compatible with the measurements within the large uncertainties.

\begin{figure} [!tb]
    \centering
    \includegraphics[width=0.49\linewidth]{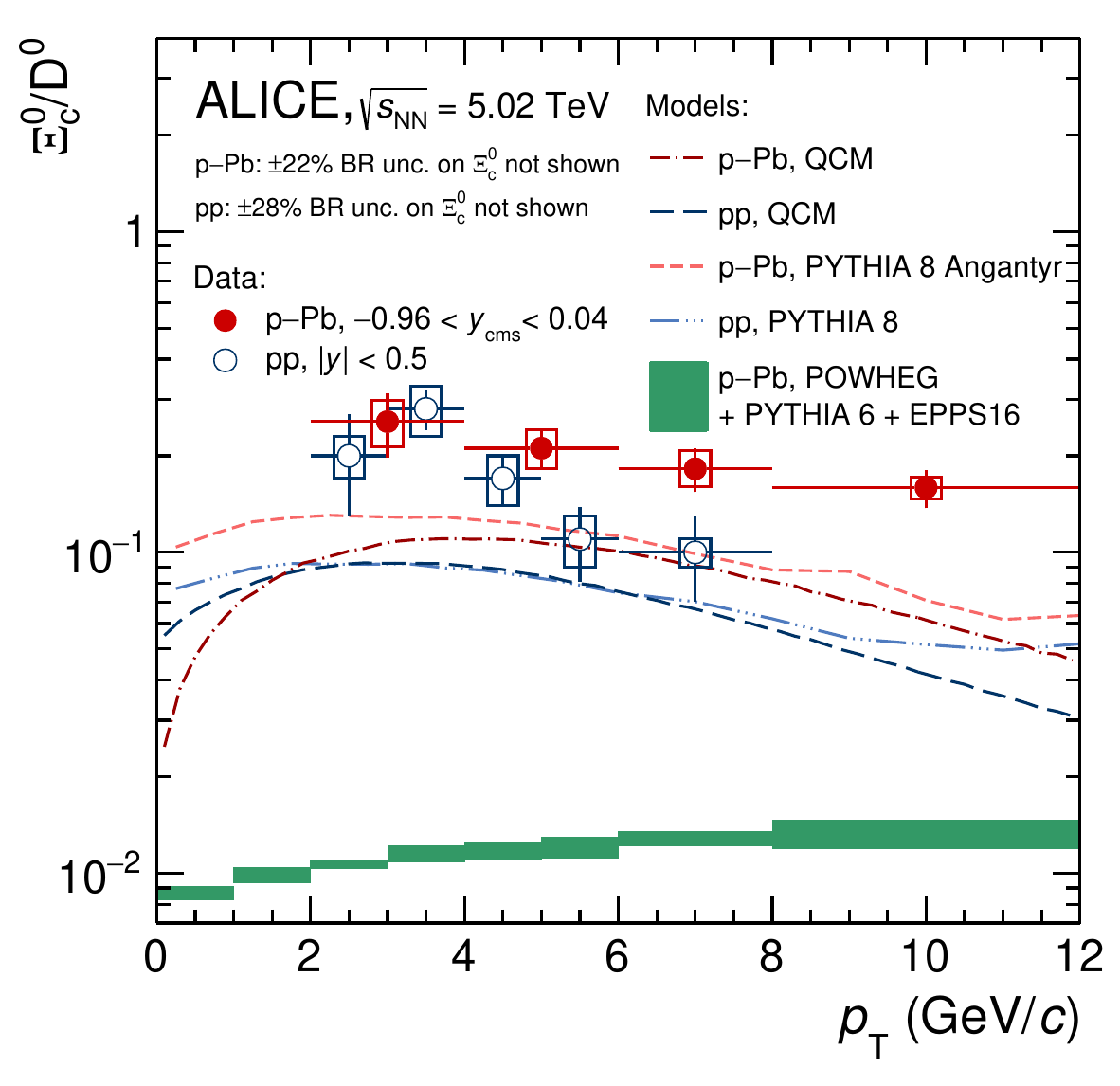}
        \includegraphics[width=0.49\linewidth]{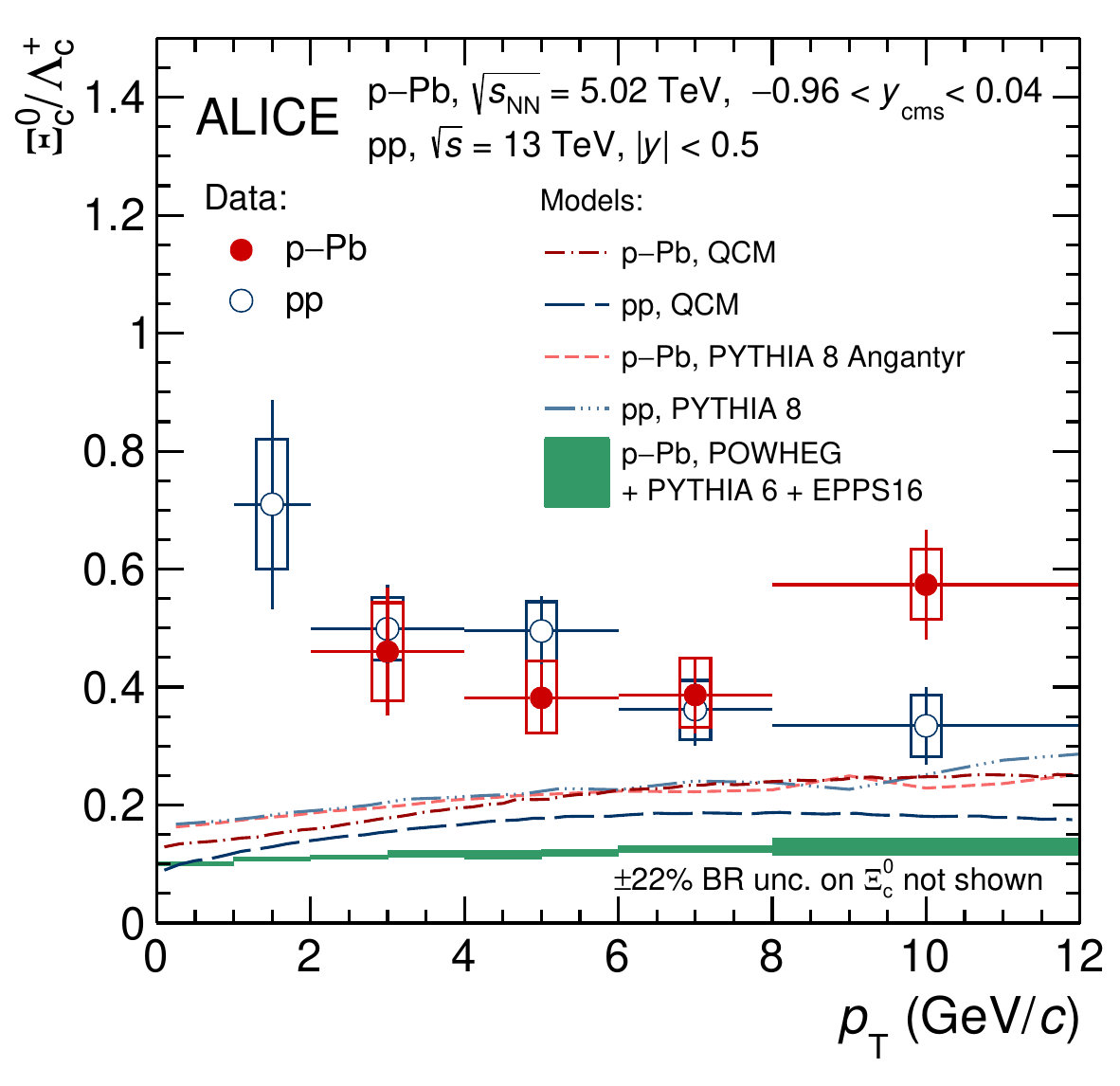}
    \caption{Left: \xicz/\DZero ratio as a function of \pt in \pPb collisions at \fivenn and in pp collisions at \five~\cite{ALICE:2021psx} compared to QCM, POWHEG + PYTHIA 6, and PYTHIA~8 Angantyr predictions. Right: \xicz/\lambdac ratio as a function of \pt in \pPb collisions at \fivenn and in \pp collisions at \thirteen~\cite{ALICE:2021bli} compared to QCM, POWHEG + PYTHIA 6, and PYTHIA~8 Angantyr calculations. The BR uncertainties for the \Lc are evaluated as the weighted average on the two decay channels~\cite{ALICE:2017thy} and are included in the systematic uncertainty box for each \pt bin.}
    \label{fig:ParticleRatio}
\end{figure}

The ratio of the production cross sections of different hadron species is sensitive to the modification of the hadronisation mechanisms in a partonic environment. The \pt-dependent $\xicz/\DZero$ baryon-to-meson yield ratio measured in pp and \pPb collisions at \fivenn~\cite{ALICE:2021psx} is reported in the left panel of Fig.~\Ref{fig:ParticleRatio}. For the measurement in \pPb collisions, the prompt \DZero cross section reported in Ref.~\cite{ALICE:2019fhe} is used.
The systematic uncertainty on the $\xicz/\DZero$ yield ratio is calculated assuming all the uncertainties of the \xicz and \DZero cross sections as uncorrelated, except for the tracking and feed-down systematic uncertainties, which partially cancel in the ratio, and the uncertainty on the luminosity which fully cancels in the ratio. The $\xicz/\DZero$ ratio in \pPb collisions hints at a slightly decreasing trend with \pt, similar to the one measured in pp collisions, albeit with large uncertainties. The theoretical predictions from QCM are also shown and they underpredict the $\xicz/\DZero$ ratios by a similar amount in both collision systems. This discrepancy is predominantly due to the low \xicz-baryon yield predicted by this model, as it can be seen by the fact that QCM undershoots the \xicz cross section (see Fig.~\ref{fig:CrossSection}) and the \xiczDz yield ratio by the same amount. 
PYTHIA~8 Angantyr calculations~\cite{Bierlich:2023okq} are also shown for both pp and p--Pb collisions. This PYTHIA~8 implementation improves the description for \xicz with respect to previous calculations~\cite{Christiansen:2015yqa} available for pp collisions. However, the model still underestimates the \xiczDz yield ratio by a similar amount as the QCM model in both pp and p--Pb collisions. One difference between the two models is that the PYTHIA~8 Angantyr calculation does not exhibit a crossing at low \pt between pp and P--Pb collisions, indicating signs of larger radial flow in p--Pb collisions, but the calculation in p--Pb collisions is larger than the pp one in the full \pt interval.
POWHEG+PYTHIA~6 calculations predict a slightly increasing trend with \pt, and undershoot the measured $\xicz/\DZero$ yield ratio by approximately a factor of 20. Given that these predictions underestimate the \DZero production only by a factor of about 2, it is reasonable to conclude that the large discrepancy in the description of the \xicz-baryon production lies primarily in the hadronisation provided by PYTHIA~6, which is tuned on $\mathrm{e^+e^-}$ collisions, rather than in the description of charm production or in an inaccurate modelling of initial state CNM effects. Furthermore, given that a difference of $1.7\sigma$ between the \xicz/\DZero yield ratio in pp and \pPb collisions is measured in the 6--8~\GeVc \pt interval, it is not possible to conclude on a possible enhancement of this baryon-over-meson yield ratio in \pPb collisions with respect to \pp collisions.

The right panel of Fig.~\ref{fig:ParticleRatio} reports the \xicz/\lambdac baryon-to-baryon yield ratio measured in \pPb collisions at \fivenn compared with the ratio measured in \pp collisions at \thirteen~\cite{ALICE:2021bli}. The two measurements are in agreement within uncertainties, and they exhibit no significant \pt dependence within the current measurement uncertainties. This suggests that there is no appreciable additional modification of the hadronisation process when moving from pp to \pPb collisions. The QCM calculations show a slightly increasing trend as a function of \pt and underestimate the measured ratio by a factor of about 2.5 (1.6) in the 2-4 (6--8) GeV/c \pt intervals. 
PYTHIA~8 Angantyr calculations in pp and \pPb collisions are compatible and show a slightly increasing trend as a function of \pt. These calculations underestimate the measured ratio by a factor of about 2.4 (1.7) in the 2-4 (6--8) GeV/c \pt intervals.  
POWHEG+PYTHIA~6 predictions are also included in the figure, underestimating the \xicz/\lambdac yield ratio by a factor of about 4 at all \pt. This reflects a similar underprediction in the production of both baryon species. However, the \xicz-baryon production appears to be underpredicted more than the \Lc-baryon one, suggesting that additional effects, such as recombination, could be considered in order to provide a better description of the results, although the large uncertainties preclude definitive conclusions. 

\begin{figure} [!tb]
    \centering
    \includegraphics[width=0.49\linewidth]{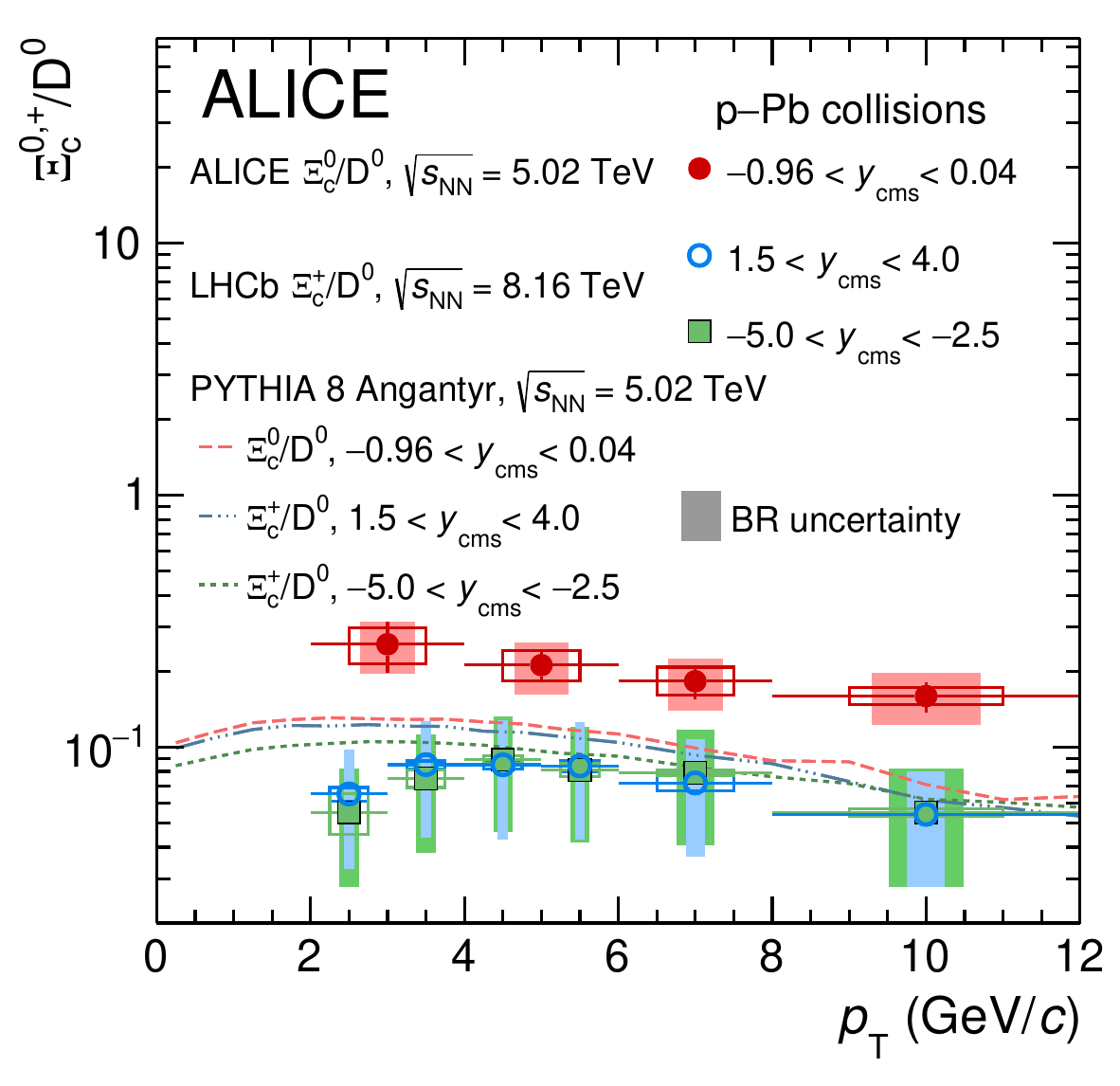}
        \includegraphics[width=0.49\linewidth]{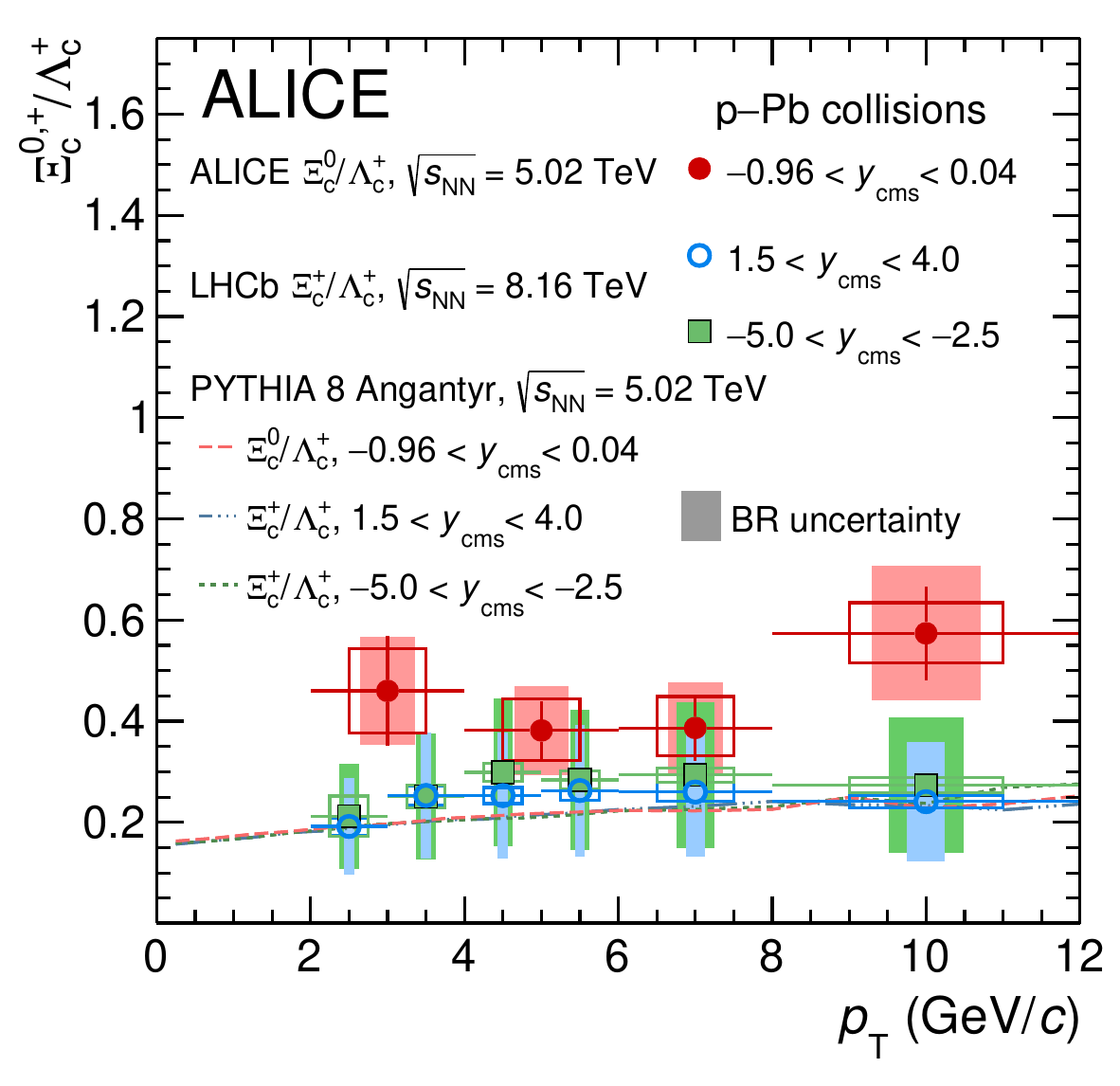}
    \caption{\xicz/\DZero (left) and \xicz/\Lc (right) ratios as a function of \pt in \pPb collisions at \fivenn measured by ALICE, compared to the \xicp/\DZero and the \xicp/\Lc ratio measured by LHCb at \eightnn~\cite{LHCb:2023cwu}, and PYTHIA~8 Angantyr predictions~\cite{Bierlich:2023okq}.}
    \label{fig:ComparisonLHCb}
\end{figure}

In Fig.~\ref{fig:ComparisonLHCb} the particle ratios measured by ALICE in the central rapidity region ($-0.96 < y_\mathrm{cms} < 0.04$) in \pPb collisions at \fivenn are compared to the $\xicp/\DZero$ (left panel) and $\xicp/\lambdac$ (right panel) yield ratios measured by LHCb at forward ($1.5 < y_\mathrm{cms} < 4.0$) and backward ($-5.0 < y_\mathrm{cms} < 2.5$) rapidities in \pPb collisions at \eightnn~\cite{LHCb:2023cwu}. A direct comparison between \xicz and \xicp results is possible because these two baryons are expected to be produced in equal amounts due to isospin symmetry and it was indeed demonstrated in \pp collisions at \thirteen that the production cross sections of these two baryon species are compatible within uncertainties~\cite{ALICE:2023sgl}. 
While the baryon-over-meson yield ratio provides an indication of a rapidity dependence, with the yield ratio at midrapidity being larger than the ones at forward and backward rapidities in the full \pt range (a difference ranging from  $1.5 \sigma$ to $2.0 \sigma$ is measured across the different \pt intervals), the baryon-over-baryon yield ratios are compatible at mid, forward, and backward rapidity within the uncertainties (a difference of $1.1 \sigma$ is measured for the 2--4 \GeVc \pt interval). Conducting measurements of baryon-over-meson and baryon-over-baryon yield ratios at the same centre-of-mass energy per nucleon pair for the same hadron reconstructed in the same decay channel will allow the removal of potential ambiguities in the interpretation of the results, helping to unravel possible rapidity dependence of the charm baryon enhancement. PYTHIA~8 Angantyr calculations, computed for p--Pb collisions for different rapidity intervals, are also shown in Fig.~\ref{fig:ComparisonLHCb}.
A small dependence on the rapidity is observed for the \xiczDz ratio. The computed ratio is largest at midrapidity and lowest at backward rapidity. This dependence might be introduced by the different parton and string densities in this asymmetric collision system. However, such a difference is not observed for the baryon-to-baryon yield ratio.  Although the baryon-to-meson and baryon-to-baryon ratios measured at midrapidity are underestimated by PYTHIA~8 Angantyr, the model agrees with the forward-rapidity measurements within uncertainties.

The visible prompt \xicz-baryon cross section is computed by integrating the \pt-differential cross section in the measured \pt region. In the integration, the systematic uncertainties were propagated considering the uncertainty due to the raw-yield extraction as fully uncorrelated and all the other sources as fully correlated between the \pt intervals. The visible \xicz-baryon cross section is

\begin{equation*}
    \mathrm{d}\sigma^{\xicz}_{\mathrm{pPb}}/\mathrm{d} y \big{|}^{(2 < \pt < 12 \text{~GeV}/c)}_{-0.96 < y < 0.04} = 8.6~\pm~1.6\,\mathrm{(stat.)}~\pm~1.3\,\mathrm{(syst.)}~\pm~1.9\,\mathrm{(BR)}~\pm~0.3\,\mathrm{(lumi.)}~\mathrm{mb}.
\end{equation*}
The \pt-integrated \xicz cross section at midrapidity is obtained by extrapolating the visible cross section to the full \pt range. The QCM spectra were chosen to perform the extrapolation, as they provide a reasonable description of the measured \pt shapes for both the production cross section and the \xiczDz baryon-to-meson yield ratio. The extrapolation factor was calculated as the ratio between the QCM predictions in the full \pt range and those in the \pt interval of the measurement. The scaling factor is 1.74. As QCM does not provide any theory uncertainties, an extrapolation uncertainty was determined using two contributions: i) the maximum difference among the extrapolation factors obtained with QCM, POWHEG+PYTHIA~6, and PYTHIA~8 Angantyr (${}^{+20.6}_{-0.0}$\%) and ii) varying the factorisation and renormalisation scales ($\mu_\mathrm{F,R}$) and the EPPS16 parameters used in the POWHEG+PYTHIA~6 calculations and determining the maximal relative differences (${}^{+23.0}_{-8.6}$\%). For the variations of $\mu_\mathrm{F,R}$, the standard variation ranges were used ($0.5 \mu_0 < \mu_\mathrm{R,F} < 2 \mu_0$, with $0.5 < \mu_{\mathrm{R}}/\mu_{\mathrm{F}} < 2$, where $\mu_0 = \sqrt{m_\mathrm{c}^2 + p_\mathrm{T}^2})$.  These contributions were added in quadrature to obtain an overall extrapolation uncertainty of ${}^{+30.9}_{-8.6}$\% on the total \xicz production cross section. The resulting \pt integrated cross section for the \xicz is

\begin{equation*}
    \mathrm{d}\sigma^{\xicz}_{\mathrm{pPb}}/\mathrm{d} y \big{|}_{-0.96 < y < 0.04} = 15.0~\pm~2.8\,\mathrm{(stat.)}~^{+2.2}_{-2.3}\,\mathrm{(syst.)}~\pm~3.3\,\mathrm{(BR)}~\pm~0.6\,\mathrm{(lumi.)}~^{+4.6}_{-1.3}\,\mathrm{(extr.)}~\mathrm{mb}. 
\end{equation*}

\section{Summary} \label{sec:summary}
The measurement of prompt \xicz-baryon production at midrapidity in \pPb collisions at \fivenn with the ALICE detector at the LHC is reported. The \pt-differential cross section for the production of prompt \xicz baryons is compared to POWHEG+PYTHIA~6 calculations. As seen in pp collisions, the results indicate a significant underestimation, by a factor of about 50, of the \xicz-baryon yield. This is mainly due to the small fraction of charm quarks hadronising into baryons in the PYTHIA~6 fragmentation stage, which is tuned on results from e$^+$e$^-$ collisions. PYTHIA~8 Angantyr predictions, with colour reconnections beyond the leading-colour approximation, include the formation of junctions that fragment into baryon. Therefore, the production of \xicz baryons is enhanced as compared to POWHEG+PYTHIA~6 calculations and the discrepancy with the measured \xicz production cross section is reduced to a factor of about 3. The QCM calculations, which implement charm-quark hadronisation via coalescence and are tuned to reproduce the \Lc/\Dzero yield ratio measured by ALICE in pp collisions at \seven, provide a better description of the measured \xicz production cross section and \xicz/\Dzero yield ratio than POWHEG+PYTHIA~6 calculations, albeit they still underestimate the \xicz-baryon yield by a factor of about 2. Neither the introduction of colour reconnections beyond the leading-colour approximation in PYTHIA~8 Angantyr, nor the coalescence mechanism as implemented in QCM quantitatively captures the \xicz enhancement relative to the \Lc. 

The \pt-differential \RpPb is larger than unity in the full \pt interval of the measurement, even though all data points are compatible with unity within the large uncertainties. The \RpPb is compatible within uncertainties with both a flat and a slightly increasing trend as the \pt increases, similar to the one observed for \Lc baryons~\cite{ALICE:2022exq}. It is underpredicted by POWHEG+PYTHIA~6 calculations with the EPPS16 parameterisation of the nuclear PDFs, suggesting that additional final-state effects might be relevant for the \xicz-baryon production in \pPb collisions, while the QCM predictions are compatible with the results within the uncertainties. PYTHIA~8 Angantyr predicitons provide a similar \pt-differential \RpPb for \xicz and \Lc baryons, exhibiting a mild increasing trend with \pt and similar magnitudes to the POWHEG+PYTHIA~6 calculations. The \xicz \RpPb is compatible with that of \Lc baryons within uncertainties, possibly hinting to a similar modification of the production of the two baryon species in \pPb collisions with respect to pp collisions. 

The \xiczDz baryon-to-meson yield ratio shows a slightly decreasing trend with increasing \pt. It is underestimated by the QCM predictions by a factor of about two. This discrepancy originates from the underestimation of the \xicz production cross section. The POWHEG+PYTHIA~6 simulations underestimate the measured ratio by a factor of 20, mainly due to the description of the charm-quark fragmentation in PYTHIA~6. The introduction of colour reconnections beyond the leading-colour approximation implemented in PYTHIA~8 Angantyr reduces the data-model discrepancy, but still leads to an underestimation of the \xiczDz yield ratio similar to that of QCM. 
The \xicz/\Lc ratio is also reported. It is compared to the results obtained in pp collisions at \thirteen. The two measurements are compatible within the uncertainties and do not present a significant \pt dependence. The measured ratios are underestimated by QCM, POWHEG+PYTHIA~6, and PYTHIA~8 Angantyr calculations.
The reported \xiczDz yield ratio is found to be larger than the measurements of \xicp/\Dzero performed by LHCb at forward rapidity, while the \xicz/\Lc and \xicp/\Lc yield ratios measured in the two rapidity intervals  are found to be compatible within uncertainties.
Lastly, the \pt-integrated \xicz production cross section is measured by extrapolating the visible cross section to the full \pt range using the \pt shape from QCM predictions. 

These measurements provide important inputs and constraints to theoretical model calculations of the hadronisation process. 
Moreover, these results will provide key inputs for the computation of the charm quark fragmentation fractions in \pPb collisions.


\newenvironment{acknowledgement}{\relax}{\relax}
\begin{acknowledgement}
\section*{Acknowledgements}

The ALICE Collaboration would like to thank all its engineers and technicians for their invaluable contributions to the construction of the experiment and the CERN accelerator teams for the outstanding performance of the LHC complex.
The ALICE Collaboration gratefully acknowledges the resources and support provided by all Grid centres and the Worldwide LHC Computing Grid (WLCG) collaboration.
The ALICE Collaboration acknowledges the following funding agencies for their support in building and running the ALICE detector:
A. I. Alikhanyan National Science Laboratory (Yerevan Physics Institute) Foundation (ANSL), State Committee of Science and World Federation of Scientists (WFS), Armenia;
Austrian Academy of Sciences, Austrian Science Fund (FWF): [M 2467-N36] and Nationalstiftung f\"{u}r Forschung, Technologie und Entwicklung, Austria;
Ministry of Communications and High Technologies, National Nuclear Research Center, Azerbaijan;
Conselho Nacional de Desenvolvimento Cient\'{\i}fico e Tecnol\'{o}gico (CNPq), Financiadora de Estudos e Projetos (Finep), Funda\c{c}\~{a}o de Amparo \`{a} Pesquisa do Estado de S\~{a}o Paulo (FAPESP) and Universidade Federal do Rio Grande do Sul (UFRGS), Brazil;
Bulgarian Ministry of Education and Science, within the National Roadmap for Research Infrastructures 2020-2027 (object CERN), Bulgaria;
Ministry of Education of China (MOEC) , Ministry of Science \& Technology of China (MSTC) and National Natural Science Foundation of China (NSFC), China;
Ministry of Science and Education and Croatian Science Foundation, Croatia;
Centro de Aplicaciones Tecnol\'{o}gicas y Desarrollo Nuclear (CEADEN), Cubaenerg\'{\i}a, Cuba;
Ministry of Education, Youth and Sports of the Czech Republic, Czech Republic;
The Danish Council for Independent Research | Natural Sciences, the VILLUM FONDEN and Danish National Research Foundation (DNRF), Denmark;
Helsinki Institute of Physics (HIP), Finland;
Commissariat \`{a} l'Energie Atomique (CEA) and Institut National de Physique Nucl\'{e}aire et de Physique des Particules (IN2P3) and Centre National de la Recherche Scientifique (CNRS), France;
Bundesministerium f\"{u}r Bildung und Forschung (BMBF) and GSI Helmholtzzentrum f\"{u}r Schwerionenforschung GmbH, Germany;
General Secretariat for Research and Technology, Ministry of Education, Research and Religions, Greece;
National Research, Development and Innovation Office, Hungary;
Department of Atomic Energy Government of India (DAE), Department of Science and Technology, Government of India (DST), University Grants Commission, Government of India (UGC) and Council of Scientific and Industrial Research (CSIR), India;
National Research and Innovation Agency - BRIN, Indonesia;
Istituto Nazionale di Fisica Nucleare (INFN), Italy;
Japanese Ministry of Education, Culture, Sports, Science and Technology (MEXT) and Japan Society for the Promotion of Science (JSPS) KAKENHI, Japan;
Consejo Nacional de Ciencia (CONACYT) y Tecnolog\'{i}a, through Fondo de Cooperaci\'{o}n Internacional en Ciencia y Tecnolog\'{i}a (FONCICYT) and Direcci\'{o}n General de Asuntos del Personal Academico (DGAPA), Mexico;
Nederlandse Organisatie voor Wetenschappelijk Onderzoek (NWO), Netherlands;
The Research Council of Norway, Norway;
Pontificia Universidad Cat\'{o}lica del Per\'{u}, Peru;
Ministry of Science and Higher Education, National Science Centre and WUT ID-UB, Poland;
Korea Institute of Science and Technology Information and National Research Foundation of Korea (NRF), Republic of Korea;
Ministry of Education and Scientific Research, Institute of Atomic Physics, Ministry of Research and Innovation and Institute of Atomic Physics and Universitatea Nationala de Stiinta si Tehnologie Politehnica Bucuresti, Romania;
Ministry of Education, Science, Research and Sport of the Slovak Republic, Slovakia;
National Research Foundation of South Africa, South Africa;
Swedish Research Council (VR) and Knut \& Alice Wallenberg Foundation (KAW), Sweden;
European Organization for Nuclear Research, Switzerland;
Suranaree University of Technology (SUT), National Science and Technology Development Agency (NSTDA) and National Science, Research and Innovation Fund (NSRF via PMU-B B05F650021), Thailand;
Turkish Energy, Nuclear and Mineral Research Agency (TENMAK), Turkey;
National Academy of  Sciences of Ukraine, Ukraine;
Science and Technology Facilities Council (STFC), United Kingdom;
National Science Foundation of the United States of America (NSF) and United States Department of Energy, Office of Nuclear Physics (DOE NP), United States of America.
In addition, individual groups or members have received support from:
Czech Science Foundation (grant no. 23-07499S), Czech Republic;
European Research Council (grant no. 950692), European Union;
ICSC - Centro Nazionale di Ricerca in High Performance Computing, Big Data and Quantum Computing, European Union - NextGenerationEU;
Academy of Finland (Center of Excellence in Quark Matter) (grant nos. 346327, 346328), Finland.
\end{acknowledgement}

\bibliographystyle{utphys} 
\bibliography{bibliography}

\newpage
\appendix

\section{The ALICE Collaboration}
\label{app:collab}
\begin{flushleft} 
\small

S.~Acharya\,\orcidlink{0000-0002-9213-5329}\,$^{\rm 127}$, 
D.~Adamov\'{a}\,\orcidlink{0000-0002-0504-7428}\,$^{\rm 86}$, 
A.~Agarwal$^{\rm 135}$, 
G.~Aglieri Rinella\,\orcidlink{0000-0002-9611-3696}\,$^{\rm 32}$, 
L.~Aglietta\,\orcidlink{0009-0003-0763-6802}\,$^{\rm 24}$, 
M.~Agnello\,\orcidlink{0000-0002-0760-5075}\,$^{\rm 29}$, 
N.~Agrawal\,\orcidlink{0000-0003-0348-9836}\,$^{\rm 25}$, 
Z.~Ahammed\,\orcidlink{0000-0001-5241-7412}\,$^{\rm 135}$, 
S.~Ahmad\,\orcidlink{0000-0003-0497-5705}\,$^{\rm 15}$, 
S.U.~Ahn\,\orcidlink{0000-0001-8847-489X}\,$^{\rm 71}$
% ...

\section*{Affiliation Notes}

$^{\rm I}$ Deceased\\
$^{\rm II}$ Also at: Max-Planck-Institut fur Physik, Munich, Germany\\
$^{\rm III}$ Also at: Italian National Agency for New Technologies, Energy and Sustainable Economic Development (ENEA), Bologna, Italy\\
$^{\rm IV}$ Also at: Dipartimento DET del Politecnico di Torino, Turin, Italy\\
$^{\rm V}$ Also at: Yildiz Technical University, Istanbul, T\"{u}rkiye\\
$^{\rm VI}$ Also at: Department of Applied Physics, Aligarh Muslim University, Aligarh, India\\
$^{\rm VII}$ Also at: Institute of Theoretical Physics, University of Wroclaw, Poland\\
$^{\rm VIII}$ Also at: An institution covered by a cooperation agreement with CERN\\

\section*{Collaboration Institutes}

$^{1}$ A.I. Alikhanyan National Science Laboratory (Yerevan Physics Institute) Foundation, Yerevan, Armenia\\
$^{2}$ AGH University of Krakow, Cracow, Poland\\
$^{3}$ Bogolyubov Institute for Theoretical Physics, National Academy of Sciences of Ukraine, Kiev, Ukraine\\
$^{4}$ Bose Institute, Department of Physics  and Centre for Astroparticle Physics and Space Science (CAPSS), Kolkata, India\\
$^{5}$ California Polytechnic State University, San Luis Obispo, California, United States\\
$^{6}$ Central China Normal University, Wuhan, China\\
$^{7}$ Centro de Aplicaciones Tecnol\'{o}gicas y Desarrollo Nuclear (CEADEN), Havana, Cuba\\
$^{8}$ Centro de Investigaci\'{o}n y de Estudios Avanzados (CINVESTAV), Mexico City and M\'{e}rida, Mexico\\
$^{9}$ Chicago State University, Chicago, Illinois, United States\\
$^{10}$ China Institute of Atomic Energy, Beijing, China\\
$^{11}$ China University of Geosciences, Wuhan, China\\
$^{12}$ Chungbuk National University, Cheongju, Republic of Korea\\
$^{13}$ Comenius University Bratislava, Faculty of Mathematics, Physics and Informatics, Bratislava, Slovak Republic\\
$^{14}$ Creighton University, Omaha, Nebraska, United States\\
$^{15}$ Department of Physics, Aligarh Muslim University, Aligarh, India\\
$^{16}$ Department of Physics, Pusan National University, Pusan, Republic of Korea\\
$^{17}$ Department of Physics, Sejong University, Seoul, Republic of Korea\\
$^{18}$ Department of Physics, University of California, Berkeley, California, United States\\
$^{19}$ Department of Physics, University of Oslo, Oslo, Norway\\
$^{20}$ Department of Physics and Technology, University of Bergen, Bergen, Norway\\
$^{21}$ Dipartimento di Fisica, Universit\`{a} di Pavia, Pavia, Italy\\
$^{22}$ Dipartimento di Fisica dell'Universit\`{a} and Sezione INFN, Cagliari, Italy\\
$^{23}$ Dipartimento di Fisica dell'Universit\`{a} and Sezione INFN, Trieste, Italy\\
$^{24}$ Dipartimento di Fisica dell'Universit\`{a} and Sezione INFN, Turin, Italy\\
$^{25}$ Dipartimento di Fisica e Astronomia dell'Universit\`{a} and Sezione INFN, Bologna, Italy\\
$^{26}$ Dipartimento di Fisica e Astronomia dell'Universit\`{a} and Sezione INFN, Catania, Italy\\
$^{27}$ Dipartimento di Fisica e Astronomia dell'Universit\`{a} and Sezione INFN, Padova, Italy\\
$^{28}$ Dipartimento di Fisica `E.R.~Caianiello' dell'Universit\`{a} and Gruppo Collegato INFN, Salerno, Italy\\
$^{29}$ Dipartimento DISAT del Politecnico and Sezione INFN, Turin, Italy\\
$^{30}$ Dipartimento di Scienze MIFT, Universit\`{a} di Messina, Messina, Italy\\
$^{31}$ Dipartimento Interateneo di Fisica `M.~Merlin' and Sezione INFN, Bari, Italy\\
$^{32}$ European Organization for Nuclear Research (CERN), Geneva, Switzerland\\
$^{33}$ Faculty of Electrical Engineering, Mechanical Engineering and Naval Architecture, University of Split, Split, Croatia\\
$^{34}$ Faculty of Engineering and Science, Western Norway University of Applied Sciences, Bergen, Norway\\
$^{35}$ Faculty of Nuclear Sciences and Physical Engineering, Czech Technical University in Prague, Prague, Czech Republic\\
$^{36}$ Faculty of Physics, Sofia University, Sofia, Bulgaria\\
$^{37}$ Faculty of Science, P.J.~\v{S}af\'{a}rik University, Ko\v{s}ice, Slovak Republic\\
$^{38}$ Frankfurt Institute for Advanced Studies, Johann Wolfgang Goethe-Universit\"{a}t Frankfurt, Frankfurt, Germany\\
$^{39}$ Fudan University, Shanghai, China\\
$^{40}$ Gangneung-Wonju National University, Gangneung, Republic of Korea\\
$^{41}$ Gauhati University, Department of Physics, Guwahati, India\\
$^{42}$ Helmholtz-Institut f\"{u}r Strahlen- und Kernphysik, Rheinische Friedrich-Wilhelms-Universit\"{a}t Bonn, Bonn, Germany\\
$^{43}$ Helsinki Institute of Physics (HIP), Helsinki, Finland\\
$^{44}$ High Energy Physics Group,  Universidad Aut\'{o}noma de Puebla, Puebla, Mexico\\
$^{45}$ Horia Hulubei National Institute of Physics and Nuclear Engineering, Bucharest, Romania\\
$^{46}$ HUN-REN Wigner Research Centre for Physics, Budapest, Hungary\\
$^{47}$ Indian Institute of Technology Bombay (IIT), Mumbai, India\\
$^{48}$ Indian Institute of Technology Indore, Indore, India\\
$^{49}$ INFN, Laboratori Nazionali di Frascati, Frascati, Italy\\
$^{50}$ INFN, Sezione di Bari, Bari, Italy\\
$^{51}$ INFN, Sezione di Bologna, Bologna, Italy\\
$^{52}$ INFN, Sezione di Cagliari, Cagliari, Italy\\
$^{53}$ INFN, Sezione di Catania, Catania, Italy\\
$^{54}$ INFN, Sezione di Padova, Padova, Italy\\
$^{55}$ INFN, Sezione di Pavia, Pavia, Italy\\
$^{56}$ INFN, Sezione di Torino, Turin, Italy\\
$^{57}$ INFN, Sezione di Trieste, Trieste, Italy\\
$^{58}$ Inha University, Incheon, Republic of Korea\\
$^{59}$ Institute for Gravitational and Subatomic Physics (GRASP), Utrecht University/Nikhef, Utrecht, Netherlands\\
$^{60}$ Institute of Experimental Physics, Slovak Academy of Sciences, Ko\v{s}ice, Slovak Republic\\
$^{61}$ Institute of Physics, Homi Bhabha National Institute, Bhubaneswar, India\\
$^{62}$ Institute of Physics of the Czech Academy of Sciences, Prague, Czech Republic\\
$^{63}$ Institute of Space Science (ISS), Bucharest, Romania\\
$^{64}$ Institut f\"{u}r Kernphysik, Johann Wolfgang Goethe-Universit\"{a}t Frankfurt, Frankfurt, Germany\\
$^{65}$ Instituto de Ciencias Nucleares, Universidad Nacional Aut\'{o}noma de M\'{e}xico, Mexico City, Mexico\\
$^{66}$ Instituto de F\'{i}sica, Universidade Federal do Rio Grande do Sul (UFRGS), Porto Alegre, Brazil\\
$^{67}$ Instituto de F\'{\i}sica, Universidad Nacional Aut\'{o}noma de M\'{e}xico, Mexico City, Mexico\\
$^{68}$ iThemba LABS, National Research Foundation, Somerset West, South Africa\\
$^{69}$ Jeonbuk National University, Jeonju, Republic of Korea\\
$^{70}$ Johann-Wolfgang-Goethe Universit\"{a}t Frankfurt Institut f\"{u}r Informatik, Fachbereich Informatik und Mathematik, Frankfurt, Germany\\
$^{71}$ Korea Institute of Science and Technology Information, Daejeon, Republic of Korea\\
$^{72}$ KTO Karatay University, Konya, Turkey\\
$^{73}$ Laboratoire de Physique Subatomique et de Cosmologie, Universit\'{e} Grenoble-Alpes, CNRS-IN2P3, Grenoble, France\\
$^{74}$ Lawrence Berkeley National Laboratory, Berkeley, California, United States\\
$^{75}$ Lund University Department of Physics, Division of Particle Physics, Lund, Sweden\\
$^{76}$ Nagasaki Institute of Applied Science, Nagasaki, Japan\\
$^{77}$ Nara Women{'}s University (NWU), Nara, Japan\\
$^{78}$ National and Kapodistrian University of Athens, School of Science, Department of Physics , Athens, Greece\\
$^{79}$ National Centre for Nuclear Research, Warsaw, Poland\\
$^{80}$ National Institute of Science Education and Research, Homi Bhabha National Institute, Jatni, India\\
$^{81}$ National Nuclear Research Center, Baku, Azerbaijan\\
$^{82}$ National Research and Innovation Agency - BRIN, Jakarta, Indonesia\\
$^{83}$ Niels Bohr Institute, University of Copenhagen, Copenhagen, Denmark\\
$^{84}$ Nikhef, National institute for subatomic physics, Amsterdam, Netherlands\\
$^{85}$ Nuclear Physics Group, STFC Daresbury Laboratory, Daresbury, United Kingdom\\
$^{86}$ Nuclear Physics Institute of the Czech Academy of Sciences, Husinec-\v{R}e\v{z}, Czech Republic\\
$^{87}$ Oak Ridge National Laboratory, Oak Ridge, Tennessee, United States\\
$^{88}$ Ohio State University, Columbus, Ohio, United States\\
$^{89}$ Physics department, Faculty of science, University of Zagreb, Zagreb, Croatia\\
$^{90}$ Physics Department, Panjab University, Chandigarh, India\\
$^{91}$ Physics Department, University of Jammu, Jammu, India\\
$^{92}$ Physics Program and International Institute for Sustainability with Knotted Chiral Meta Matter (SKCM2), Hiroshima University, Hiroshima, Japan\\
$^{93}$ Physikalisches Institut, Eberhard-Karls-Universit\"{a}t T\"{u}bingen, T\"{u}bingen, Germany\\
$^{94}$ Physikalisches Institut, Ruprecht-Karls-Universit\"{a}t Heidelberg, Heidelberg, Germany\\
$^{95}$ Physik Department, Technische Universit\"{a}t M\"{u}nchen, Munich, Germany\\
$^{96}$ Politecnico di Bari and Sezione INFN, Bari, Italy\\
$^{97}$ Research Division and ExtreMe Matter Institute EMMI, GSI Helmholtzzentrum f\"ur Schwerionenforschung GmbH, Darmstadt, Germany\\
$^{98}$ Saga University, Saga, Japan\\
$^{99}$ Saha Institute of Nuclear Physics, Homi Bhabha National Institute, Kolkata, India\\
$^{100}$ School of Physics and Astronomy, University of Birmingham, Birmingham, United Kingdom\\
$^{101}$ Secci\'{o}n F\'{\i}sica, Departamento de Ciencias, Pontificia Universidad Cat\'{o}lica del Per\'{u}, Lima, Peru\\
$^{102}$ Stefan Meyer Institut f\"{u}r Subatomare Physik (SMI), Vienna, Austria\\
$^{103}$ SUBATECH, IMT Atlantique, Nantes Universit\'{e}, CNRS-IN2P3, Nantes, France\\
$^{104}$ Sungkyunkwan University, Suwon City, Republic of Korea\\
$^{105}$ Suranaree University of Technology, Nakhon Ratchasima, Thailand\\
$^{106}$ Technical University of Ko\v{s}ice, Ko\v{s}ice, Slovak Republic\\
$^{107}$ The Henryk Niewodniczanski Institute of Nuclear Physics, Polish Academy of Sciences, Cracow, Poland\\
$^{108}$ The University of Texas at Austin, Austin, Texas, United States\\
$^{109}$ Universidad Aut\'{o}noma de Sinaloa, Culiac\'{a}n, Mexico\\
$^{110}$ Universidade de S\~{a}o Paulo (USP), S\~{a}o Paulo, Brazil\\
$^{111}$ Universidade Estadual de Campinas (UNICAMP), Campinas, Brazil\\
$^{112}$ Universidade Federal do ABC, Santo Andre, Brazil\\
$^{113}$ Universitatea Nationala de Stiinta si Tehnologie Politehnica Bucuresti, Bucharest, Romania\\
$^{114}$ University of Cape Town, Cape Town, South Africa\\
$^{115}$ University of Derby, Derby, United Kingdom\\
$^{116}$ University of Houston, Houston, Texas, United States\\
$^{117}$ University of Jyv\"{a}skyl\"{a}, Jyv\"{a}skyl\"{a}, Finland\\
$^{118}$ University of Kansas, Lawrence, Kansas, United States\\
$^{119}$ University of Liverpool, Liverpool, United Kingdom\\
$^{120}$ University of Science and Technology of China, Hefei, China\\
$^{121}$ University of South-Eastern Norway, Kongsberg, Norway\\
$^{122}$ University of Tennessee, Knoxville, Tennessee, United States\\
$^{123}$ University of the Witwatersrand, Johannesburg, South Africa\\
$^{124}$ University of Tokyo, Tokyo, Japan\\
$^{125}$ University of Tsukuba, Tsukuba, Japan\\
$^{126}$ Universit\"{a}t M\"{u}nster, Institut f\"{u}r Kernphysik, M\"{u}nster, Germany\\
$^{127}$ Universit\'{e} Clermont Auvergne, CNRS/IN2P3, LPC, Clermont-Ferrand, France\\
$^{128}$ Universit\'{e} de Lyon, CNRS/IN2P3, Institut de Physique des 2 Infinis de Lyon, Lyon, France\\
$^{129}$ Universit\'{e} de Strasbourg, CNRS, IPHC UMR 7178, F-67000 Strasbourg, France, Strasbourg, France\\
$^{130}$ Universit\'{e} Paris-Saclay, Centre d'Etudes de Saclay (CEA), IRFU, D\'{e}partment de Physique Nucl\'{e}aire (DPhN), Saclay, France\\
$^{131}$ Universit\'{e}  Paris-Saclay, CNRS/IN2P3, IJCLab, Orsay, France\\
$^{132}$ Universit\`{a} degli Studi di Foggia, Foggia, Italy\\
$^{133}$ Universit\`{a} del Piemonte Orientale, Vercelli, Italy\\
$^{134}$ Universit\`{a} di Brescia, Brescia, Italy\\
$^{135}$ Variable Energy Cyclotron Centre, Homi Bhabha National Institute, Kolkata, India\\
$^{136}$ Warsaw University of Technology, Warsaw, Poland\\
$^{137}$ Wayne State University, Detroit, Michigan, United States\\
$^{138}$ Yale University, New Haven, Connecticut, United States\\
$^{139}$ Yonsei University, Seoul, Republic of Korea\\
$^{140}$  Zentrum  f\"{u}r Technologie und Transfer (ZTT), Worms, Germany\\
$^{141}$ Affiliated with an institute covered by a cooperation agreement with CERN\\
$^{142}$ Affiliated with an international laboratory covered by a cooperation agreement with CERN.\\

\end{flushleft} 

\end{document}